\documentclass[twocolumn,prd,nofootinbib,aps,prl,floats,floatfix,amsmath,amssymb,longbibliography,secnumarab
ic]{revtex4-1} %
\usepackage[final]{graphicx}
\usepackage{hyperref}
\usepackage{amsmath}
\usepackage{bbm}
\usepackage{amsfonts}
\usepackage{amssymb}
\usepackage{latexsym}
\usepackage{graphicx}
\usepackage[english]{babel}
\usepackage{multirow}
\usepackage{float}
\usepackage{url}
\usepackage{slashed}
\usepackage{xcolor} 
\usepackage[utf8]{inputenc}
\usepackage{verbatim}
\usepackage{stackengine}
\usepackage{cancel}
\usepackage{accents}
\usepackage{array}

\def\sfrac#1#2{{\textstyle{#1\over #2}}}
\newcommand{\lp}{\left(}
\newcommand{\rp}{\right)}
\newcommand{\lbr}{\left[}
\newcommand{\rbr}{\right]}
\newcommand{\lb}{\left\lbrace}

\newcommand{\lc}{\left[}
\newcommand{\rc}{\right]}
\newcommand{\be}{\begin{equation}}
\newcommand{\ee}{\end{equation}}
\newcommand{\ba}{\begin{array}}
\newcommand{\ea}{\end{array}}
\newcommand{\bea}{\begin{eqnarray}}
\newcommand{\eea}{\end{eqnarray}}

\newcommand{\jc}[1]{{\color{red}{#1}}}

\newcommand{\nn}{\nonumber}

\newcommand{\eq}{\mathrm{eq}}
\newcommand{\out}{\mathrm{out}}
\newcommand{\plasma}{\mathrm{pl}}

\renewcommand{\bar}[1]{\accentset{\rule{.4em}{.5pt}}{#1}}

\newcommand{\bl}[1]{\color{magenta}BL: #1 \color{black}}

\begin{document}

\title{First principles determination of bubble wall velocity}
\author{Benoit Laurent}
\author{James M.\ Cline}
\affiliation{McGill University, Department of Physics, 3600 University St.,
Montr\'eal, QC H3A2T8 Canada}

\begin{abstract}

The terminal wall velocity of a first-order phase transition bubble can be calculated from a set of fluid equations describing the scalar fields and the plasma's state. We rederive these equations from the energy-momentum tensor conservation and the Boltzmann equation, without linearizing in the background temperature and fluid velocity. The resulting equations have a finite solution for any wall velocity. We propose a spectral method to integrate the Boltzmann equation, which is simple, efficient and accurate. As an example, we apply this new methodology to the singlet scalar extension of the standard model. We find that all solutions are naturally categorized as deflagrations ($v_w\sim c_s$) or ultrarelativistic detonations ($\gamma_w\gtrsim10$). Furthermore, the contributions from out-of-equilibrium effects are, most of the time, subdominant. Finally, we use these results to propose several approximation schemes with increasing levels of complexity and accuracy. They can be used to considerably simplify the methodology while correctly describing the qualitative behavior of the bubble wall.

\end{abstract}
\maketitle

\section{Introduction}

First-order phase transitions (FOPTs) can produce striking cosmological signatures that may provide a window into high-energy physics and the history of the early universe. Recently, the prospect of probing these signals for the first time through the upcoming space-based gravitational wave (GW) detector LISA \cite{Audley:2017drz,Caprini:2015zlo,Caprini:2019egz}, DECIGO \cite{Seto:2001qf,Sato:2017dkf} and BBO \cite{Crowder:2005nr,Harry:2006fi} 
has stimulated a great interest in models that predict strong FOPTs \cite{Grojean:2006bp,Dorsch:2016nrg,Vaskonen:2016yiu,Huang:2016cjm,Demidov:2017lzf,Chala:2018ari,Ahriche:2018rao,Alves:2018jsw,Wang:2019pet,Mohamadnejad:2019vzg,Caldwell:2022qsj,Banta:2022rwg}.  Such FOPTs can  also potentially provide the departure from equilibrium needed for baryogenesis and thereby explain the origin of the baryon asymmetry of the universe. One of the popular  scenarios  is electroweak baryogenesis \cite{Bochkarev:1990fx,Cohen:1990it,Cohen:1990py,Turok:1990zg} because of its testability in collider experiments, in realizations involving simple extensions of the Standard Model (SM) \cite{Anderson:1991zb,Dine:1991ck,McDonald:1993ey,Choi:1993cv,
Carena:2004ha,Cline:2011mm,Cline:2012hg,Cheung:2012pg,Inoue:2015pza,Dorsch:2016nrg,Cline:2017qpe,Akula:2017yfr,deVries:2017ncy,Bruggisser:2018mrt,Carena:2018cjh,Ellis:2019flb,Carena:2019xrr,Bell:2019mbn,Cline:2021iff,Cline:2021dkf,Angelescu:2021pcd,Barrow:2022gsu}.

To accurately predict  the cosmological signature of a FOPT, it is important to understand its dynamics in some detail. An essential quantity, which is also notoriously challenging to compute, is the bubble wall terminal velocity, $v_w$. Several methods have been developed to determine it,  as described below, which all essentially consist in requiring that the driving force on the wall be equal to the friction, or \textit{backreaction}, force from the plasma. The main differences between the methods is in how the plasma's distribution functions are represented and calculated. 

A common strategy is to assume that all the species in the plasma are in local thermal equilibrium at the same temperature and fluid velocity \cite{Konstandin:2010dm,BarrosoMancha:2020fay,Balaji:2020yrx,Ai:2021kak}. These two thermodynamic quantities can then be computed from a set of hydrodynamic equations derived from the conservation of the energy-momentum tensor (EMT). A more general approach is to allow the heavy species to be out of equilibrium, with the remaining degrees of freedom forming a \textit{background} plasma in equilibrium \cite{Moore:1995si,Moore:1995ua,Konstandin:2014zta,Kozaczuk:2015owa,Laurent:2020gpg,Dorsch:2021nje,DeCurtis:2022hlx}. These distribution functions are usually computed from a set of Boltzmann equations. The procedure introduced in this work uses some aspects of both of these methods.

The standard formalism  to solve the Boltzmann equations consists in using an {ansatz} for the distribution function.
In the seminal Refs.\ \cite{Moore:1995si,Moore:1995ua}, the out-of-equilibrium species were only allowed to deviate from the background distribution function by a perturbation of the temperature, velocity and chemical potential. The  Boltzmann equation  was then linearized in these perturbations and truncated by taking three linearly independent moments. This procedure yields a set of three moment equations that can be solved for the perturbations.

Strikingly, these equations suffer from a singularity when the wall velocity $v_w$ reaches the speed of sound $c_s\approx 1/\sqrt{3}$, which makes the background perturbations diverge. This prevents an accurate description of fast-moving walls, which has given rise to two schools of thought. The first one considers that this singularity of the moment equations is unphysical and is merely an artifact of an unsuitable representation of the full Boltzmann equation \cite{Cline:2020jre,Laurent:2020gpg}. More specifically, it was argued that the singularity is a consequence of a bad choice of moments or {ansatz}, and instead a truncation scheme for the moment expansion was
proposed, which yields finite solutions even at $v_w=c_s$. On the other hand, Refs.\ \cite{Dorsch:2021nje,Dorsch:2021ubz} argued that the singularity is caused by the linearization of the Boltzmann equation, and they proposed a generalized {ansatz} for the out-of-equilibrium distribution functions to more accurately represent the full Boltzmann equation, while interpreting that the singularity has a  physical origin: a ``sonic boom.''

In this work, we advocate a middle ground, by maintaining that the singularity of the moment expansion is unphysical and can be induced by an overly restrictive {ansatz}, while recognizing that the linearization of the background perturbations is the root of the problem, as was argued in Ref.\ \cite{Dorsch:2021nje}.  (In contrast, the out-of-equilibrium perturbations can be safely linearized.)  Based on this interpretation, we propose a solution to the problem of supersonic walls by solving nonlinearly for the background perturbations.   Then the only danger in choosing a particular ansatz is that it may not give a very accurate representation of the exact solution.  Our method will also avoid this pitfall by expanding the perturbations in a large enough basis of orthogonal polynomials so that convergence to the exact solution is achieved.

We start in Sec.\ \ref{sect:fluid} by rederiving the fluid equations used to compute the wall velocity. Our methodology differs from the standard approach by linearizing only when necessary to obtain equations that are numerically tractable. This strategy leads to a set of Boltzmann equations for each heavy out-of-equilibrium species, in which we only linearize the collision operators, and a set of hydrodynamic equations that describe the background fluid in local thermal equilibrium. We do not linearize the latter, which yields a finite solution for every wall velocity.

Then, we propose in Sec.\ \ref{sect:spectral} a novel spectral method to solve the Boltzmann equations. It uses a spectral {ansatz} that can be as general as needed, and yields a high accuracy with great efficiency. Moreover, we believe it simpler to implement than the standard moment expansion. In Sec.\ \ref{sect:xSM}, we apply this new methodology to a benchmark model: the $Z_2$-symmetric singlet scalar extension of the SM. We study its consequences for the wall velocity and shape, and reassess the importance of the out-of-equilibrium contributions. We use these results in Sec.\ \ref{sect:approx} to motivate several approximation schemes with increasing levels of complexity and accuracy. They allow to correctly describe the qualitative behavior of the wall dynamics, while substantially simplifying the fluid equations. Finally, conclusions are given in Sec.\ \ref{sect:concl}.

\section{Fluid equations}
\label{sect:fluid}
The wall terminal velocity and shape can be determined by solving a set of fluid equations describing the evolution of the plasma and the scalar fields. These equations consist of the EMT conservation and a set of Boltzmann equations describing the plasma's departure from equilibrium. In general, a static solution corresponding to the terminal state of the wall can only be found for discrete values of $v_w$. 

In this section, we derive these fluid equations in a completely Lorentz invariant way. They form a highly nontrivial nonlinear eigenvalue problem that must be solved for the wall velocity. 

\subsection{Boltzmann equation}\label{sec:BE}
The relativistic Boltzmann equation that describes the evolution of the distribution functions $f_i(p^\mu,x^\mu)$ takes the form
\be\label{eq:BE}
(p^\mu\partial_\mu+m_i F_i^\mu\partial_{p^\mu})f_i(p^\mu,x^\mu)=-\mathcal{C}_i[f_j]\,,
\ee
where  $m_i F_i^\mu=\partial^\mu(m_i^2)/2$ is the $CP$-even force applied on the particles by the moving wall\footnote{A more general expression would also include a $CP$-odd force, which plays an important role in baryogenesis.} and $\mathcal{C}[f_j]$ is a nonlinear collision integral. To keep Lorentz invariance manifest, the 4-momentum $p^\mu$ is set to its on-shell value ($p^0=\sqrt{m^2+|\mathbf{p}|^2}$) only after the derivatives of $f_i$ are taken, so that one can assume $\partial_\mu p^\nu=0$ and $\partial_{p^\mu}p^\nu=\delta_\mu^\nu$.

If the bubble radius is much larger than its width, the wall can be approximated to be planar and moving in the $+z$ direction. In that case, translational invariance in the plane parallel to the wall implies that none of the quantities in Eq.\ (\ref{eq:BE}) depend on the $x$ and $y$ coordinates. Furthermore, they should only depend on the distance from the wall $\xi\equiv -{\bar u}_w^\mu x_\mu=\gamma_w(z-v_w t)$, where ${\bar u}_w^\mu=\gamma_w(v_w,0,0,1)$ is the 4-velocity perpendicular to the wall 4-velocity $u_w^\mu=\gamma_w(1,0,0,v_w)$. This allows us to write the derivative with respect to the coordinates as
\be\label{eq:derivative}
\partial^\mu = \partial^\mu\xi\partial_\xi = -{\bar u}_w^\mu\partial_\xi\,.
\ee

In general, the distribution functions can be written as the sum of an equilibrium distribution 
\be
f_i^\eq(p^\mu,\xi)=
{1\over \exp[p_\mu u_\plasma^\mu(\xi) /T(\xi)]\pm1}
\ee
and some function describing the deviation from equilibrium $\delta\! f_i(\Vec{p},\xi)$:
\be\label{eq:distrFunc}
f_i(p^\mu,\xi)=f_i^\eq(p^\mu,\xi)+\delta\! f_i(p^\mu,\xi)\,.
\ee
In this parametrization, all the species share the same local temperature $T(\xi)$ and local plasma 4-velocity $u_\plasma^\mu(\xi)\equiv \gamma_\plasma(\xi)(1,0,0,v_\plasma(\xi))$, which can both depend on $\xi$. We use the convention where $v_\plasma\leq0$ and $v_w\geq0$, which corresponds to the wall moving in the positive $z$ direction.

{In previous studies, the position-dependent temperature was generally written as $T(\xi)={\bar T}+\delta T_\mathrm{bg}(\xi)$ (and similarly for $v_\plasma(\xi)$), where ${\bar T}$ is an arbitrary constant temperature, and $\delta T_\mathrm{bg}$ is a small temperature deviation of the ``background" species (the light species in local thermal equilibrium). This deviation was found by solving a Boltzmann equation similar to Eq.\ (\ref{eq:BE}) and linearized in $\delta T_\mathrm{bg}$. However, Ref.\ \cite{Dorsch:2021nje} has argued that this linearization leads to a singularity of the fluid equations which makes the fluid temperature and velocity diverge when the wall velocity approaches the speed of sound. Fortunately for us, it is not necessary to linearize the Boltzmann in the background perturbations to make it numerically tractable. It is sufficient to linearize only the collision operator $\mathcal{C}_i[\delta\! f_j]$ in terms of $\delta\! f_j$, which does not induce any singularity in the equations.\footnote{The failure of linearization in $\delta T_{\rm bg}$ and $\delta v_{\rm bg}$ occurs not because these quantities ever become large, but rather because the Liouville operator of the Boltzmann equation becomes singular unless higher order terms are retained \cite{Dorsch:2021nje}.} 

One might wonder whence this fundamental difference between the background and out-of-equilibrium perturbations arises, that only the former causes a singularity when linearized. There are two main reasons for the differing behaviors: the background perturbations are constrained by EMT conservation and they are the only perturbations undamped by $\mathcal{C}_i[f_j]$. The constraints from EMT conservation lead to a set of hydrodynamic equations derived in the next subsection. Ref.\ \cite{Dorsch:2021nje} has showed that, when linearized, these equations are singular at $v_w=c_s$ if local thermal equilibrium is assumed. One could try to solve the problem by relaxing this last assumption, by allowing the background fluid to be out of equilibrium. While this might work close to the wall, it is doomed to fail away from it since all these out-of-equilibrium perturbations are exponentially damped by $\mathcal{C}_i[f_j]$, which ensures that the plasma is in local thermal equilibrium at infinity.

Therefore, the only way to remove the unphysical singularity is to keep the full nonlinear dependence on $T(\xi)$ and $v_\plasma(\xi)$. We determine these directly from the hydrodynamic equations,\footnote{This method has already been used in Refs.\ \cite{Balaji:2020yrx,Ai:2021kak} for a plasma in local thermal equilibrium. We here extend their derivation to include out-of-equilibrium perturbations.} which will be derived in the next subsection.}

Substituting Eqs.\ (\ref{eq:derivative}) and (\ref{eq:distrFunc}) into Eq.\ (\ref{eq:BE}), one obtains
\be\label{eq:BESimplified}
\lp\mathcal{P}_{\!w}\partial_\xi-\frac{1}{2}\partial_\xi(m_i^2)\bar{u}_w^\mu\partial_{p^\mu}\rp\delta\! f_i=-\mathcal{C}_i^\mathrm{lin}[\delta\! f_j]+\mathcal{S}_i\,,
\ee
where we introduce the notation $\mathcal{E}_{a}\equiv p_\mu u_a^\mu$ and $\mathcal{P}_{\!a}\equiv -p_\mu {\bar u}_a^\mu$ with $a\in\{\plasma,w\}$, which are Lorentz invariant quantities corresponding respectively to the energy and $z$-momentum in the plasma ($\plasma$) and wall ($w$) frames. $\mathcal{C}^\mathrm{lin}[\delta\! f]$ is the linearized collision operator, which can safely be used instead of $\mathcal{C}[\delta\! f]$ since $\delta\! f$ is not subject to the same constraint as the background perturbations, that caused the singularity. Moreover, the deviation from equilibrium is expected to be small, so neglecting the higher order terms should be a reasonable approximation. See Appendix \ref{app:coll} for more details about $\mathcal{C}^\mathrm{lin}[\delta\! f]$. The source term  in Eq.\ (\ref{eq:BESimplified}) is
\bea\label{eq:source}
\mathcal{S}_i &=& -\lp\mathcal{P}_{\!w}\partial_\xi-\frac{1}{2}\partial_\xi(m_i^2)\bar{u}_w^\mu\partial_{p^\mu}\rp f_i^\eq \\
&=& f_i'\frac{\mathcal{P}_{\!w}}{T}\left[\gamma_\plasma^2\mathcal{P}_{\!\plasma} \partial_\xi v_\plasma+\mathcal{E}_\plasma\frac{\partial_\xi T}{T}\right] 
+\frac{1}{2} \partial_\xi(m_i^2)\bar{u}_{w,\mu}u_\plasma^\mu \frac{f_i'}{T}\,,\nn
\eea
with
\be
f_i'\equiv -\frac{e^{\mathcal{E}_\plasma/T}}{(e^{\mathcal{E}_\plasma/T}\pm1)^2}\,.
\ee

\subsection{Hydrodynamic equations}

The starting point to derive the hydrodynamic equations and the scalar fields' equations of motion is the conservation of stress-energy. The EMT of the vacuum contribution from $N$ scalar fields $\phi_i$ is 
\be
T_\phi^{\mu\nu}=\partial^\mu\phi_i\partial^\nu\phi_i-\eta^{\mu\nu}\left[\frac{1}{2}\partial_\alpha\phi_i\partial^\alpha\phi_i-V_0(\phi_j)\right]\,,
\ee
where repeated indices are summed over, and $V_0(\phi_j)$ is the scalar fields' vacuum potential, including loop corrections. One also needs to include the contribution from the plasma, which can be written in terms of the distribution functions $f_i$ as
\be
T_\plasma^{\mu\nu}=\sum_i\int\frac{d^3p}{(2\pi)^3E_i}\,p^\mu p^\nu f_i(p^\mu,\xi)\,,
\ee
where the sum is over all the plasma's degrees of freedom.

One can substitute the {ansatz} (\ref{eq:distrFunc}) to express $T_\plasma^{\mu\nu}$ as a sum of equilibrium and out-of-equilibrium contributions:
\bea
T_\plasma^{\mu\nu}&=&T_\eq^{\mu\nu}+T_\out^{\mu\nu}\nn\\
T_\eq^{\mu\nu} &=& \sum_i\int\frac{d^3p}{(2\pi)^3E_i}\,p^\mu p^\nu f_i^\eq(p^\mu,\xi)\nn\\
T_\out^{\mu\nu} &=& \sum_i\int\frac{d^3p}{(2\pi)^3E_i}\,p^\mu p^\nu\, \delta\! f_i(p^\mu,\xi)\,.
\eea
Using Lorentz covariance, one can write $T_\eq^{\mu\nu}$ as
\be
T_\eq^{\mu\nu} = w\, u_\plasma^\mu u_\plasma^\nu - p\, \eta^{\mu\nu}\,,
\label{Teqeq}
\ee
where the thermal pressure $p$ and enthalpy $w$ are
\bea
p&=&\pm T\sum_i \int\frac{d^3p}{(2\pi)^3}\log\left[1\pm\exp(-E_i/T)\right]\,,\nn\\
w &=& T\frac{\partial p}{\partial T}\,,
\eea
with $E_i$ defined in the plasma rest frame and the $+/-$ signs denoting fermions/bosons. 

Similarly to Eq.\ (\ref{Teqeq}), one can use Lorentz covariance and the symmetry of the EMT under $\mu\leftrightarrow\nu$ to express $T_\out^{\mu\nu}$ as a linear combination of the tensors $\eta^{\mu\nu}$, $u_\plasma^\mu u_\plasma^\nu$, ${\bar u}_\plasma^\mu {\bar u}_\plasma^\nu$ and $(u_\plasma^\mu {\bar u}_\plasma^\nu+{\bar u}_\plasma^\mu u_\plasma^\nu)$:
\be
T_\out^{\mu\nu} = T_{\out,\eta}\,\eta^{\mu\nu}+T_{\out,u}^{\mu\nu}\,,
\ee
with
\bea
T_{\out,\eta} &=& \frac{1}{2}\sum_i (m_i^2 \Delta_{00}^i+\Delta_{02}^i-\Delta_{20}^i),\nn \\
T_{\out,u}^{\mu\nu} &=& \frac{1}{2}\sum_i\Big[(3\Delta_{20}^i-\Delta_{02}^i-m_i^2 \Delta_{00}^i)u_\plasma^\mu u_\plasma^\nu\nn\\
&&+(3\Delta_{02}^i-\Delta_{20}^i+m_i^2 \Delta_{00}^i){\bar u}_\plasma^\mu {\bar u}_\plasma^\nu \nn\\
&&+2\Delta_{11}^i(u_\plasma^\mu {\bar u}_\plasma^\nu+{\bar u}_\plasma^\mu u_\plasma^\nu)\Big]\,,
\eea
and
\be\label{eq:Delta}
\Delta_{mn}^i(\xi)=\int\frac{d^3p}{(2\pi)^3E}\, \mathcal{E}_\plasma^m \mathcal{P}_\plasma^n\, \delta\! f_i(p^\mu,\xi)\,.
\ee

Conservation of the EMT is then given by
\bea\label{eq:EMTcons}
0&=&\partial_\mu T^{\mu\nu}=\partial_\mu (T_\phi^{\mu\nu}+T_\eq^{\mu\nu}+T_\out^{\mu\nu})\nn\\
&=&\partial^\nu\phi_i\left[ \partial^2\phi_i+\frac{\partial V_T(\phi_j,T)}{\partial \phi_i}+\frac{\partial T_{\out,\eta}}{\partial \phi_i}\right]-\partial^\nu T\frac{\partial p}{\partial T}\nn\\
&&+\left(\partial^\nu T_{\out,\eta}\right)_\phi+\partial_\mu(w\, u_\plasma^\mu u_\plasma^\nu+T_{\out,u}^{\mu\nu})\,,
\eea
where $V_T\equiv V_0-p$ and the subscript $\phi$ indicates that the scalar fields $\phi_i$ are kept constant. We recognize the quantity in brackets as the equation of motion for the scalar fields, which must vanish independently of the rest of Eq.\ (\ref{eq:EMTcons}):
\be\label{eq:EOM}
\partial^2\phi_i+\frac{\partial V_T(\phi_j,T)}{\partial \phi_i}+\sum_j\frac{\partial (m_j^2)}{\partial \phi_i}\frac{\Delta_{00}^j}{2}=0\,,
\ee
where we assume that the only dependence of $T_{\out,\eta}$ on $\phi_i$ comes from the explicit factor of $m_j^2$ multiplying $\Delta_{00}^j$.

To derive the last two equations, that describe the evolution of $T(\xi)$ and $v_\plasma(\xi)$, it is convenient to write Eq.\ (\ref{eq:EMTcons}) in the frame of the wall, where all the quantities depend only on the $z$ coordinate. Conservation of the EMT can then be written as
\bea\label{eq:EMT}
T^{30} &=& w\gamma_\plasma^2v_\plasma+T_{\out}^{30}=c_1\,,\\
T^{33} &=& \frac{1}{2}(\partial_z\phi_i)^2-V_T(\phi_j,T)
+ w\gamma_\plasma^2v_\plasma^2+T_{\out}^{33}=c_2\,,\nn
\eea
where $c_1$ and $c_2$ are constants that depend on $T_-$ and $v_-$ (or alternatively on $T_+$ and $v_+$), which denote the fluid temperature and velocity at $z\rightarrow \mp\infty$. These can be computed as functions of $v_w$ using the method described in Ref.\ \cite{Espinosa:2010hh}, which is summarized in Appendix \ref{app:BC}.

In practice, one can directly solve the first line of Eq.\ (\ref{eq:EMT}) for $v_\plasma$:
\be
v_\plasma = \frac{-w+\sqrt{4s_1^2+w^2}}{2s_1}\,,
\ee
with $s_1=c_1-T_\out^{30}$. Substituting $v_{\rm pl}$ into
the equation for $T^{33}$ yields
\be\label{eq:EMTsimplified}
\frac{1}{2}(\partial_z\phi_i)^2-V_T-\frac{1}{2}w+\frac{1}{2}\sqrt{4s_1^2+w^2}-s_2=0\,,
\ee
where  $s_2=c_2-T_\out^{33}$; this can be  solved numerically for the temperature as a function of the scalar fields.

In addition to Eqs.\ (\ref{eq:EOM}) and (\ref{eq:EMTsimplified}), one must impose boundary conditions on the scalar fields to insure that they start and end in the false and true vacua, respectively:
\be\label{eq:BC}
\phi_i(z\rightarrow\pm\infty) = \phi_i^\pm,
\ee
where $\phi_i^\pm$ satisfy
\be
0=\left. \frac{\partial V_T}{\partial \phi_i}\right\vert_{\phi_j=\phi_j^\pm,T=T_\pm}\,.
\ee
However, the boundary conditions (\ref{eq:BC}) are  insufficient for specifying a unique solution. Effectively, all the hydrodynamic equations and boundary conditions derived so far are invariant under translations along the $z$ axis. This implies that there exists a continuous family of solutions related to one another via the substitution $z\rightarrow z+a,\ a\in\mathbb{R}$. To remove this degeneracy, one must give an additional boundary condition that specifies the position of the wall. We choose here to impose
\be\label{eq:BC0}
\phi_1(z=0) = \frac{\phi_1^-+\phi_1^+}{2}\,.
\ee

The resulting system of equations naively appears to be overconstrained, as there are now more boundary conditions than the number of differential equations. The additional constraint serves to determine the unknown wall velocity $v_w$, 
which plays the role of a nonlinear eigenvalue for the system of equations (\ref{eq:EOM},\ref{eq:EMT},\ref{eq:BC},\ref{eq:BC0}). Formally, this can be handled by promoting $v_w$ to an undetermined function $v_w(z)$ with the differential equation
\be
\partial_z v_w(z) = 0\,,
\ee
which, of course, enforces that $v_w$ is a constant.

\section{Spectral solution of the Boltzmann equation}
\label{sect:spectral}
The Boltzmann equation (\ref{eq:BESimplified}) is notoriously difficult to solve, being a partial integro-differential equation involving the nonlocal collision operator $\mathcal{C}[f]$, which can be laborious to compute. We present here a spectral method that can efficiently deal with such difficulties.
The basic idea behind the spectral method is to expand the unknown functions in series of orthogonal polynomials. Doing this transforms the differential equation into an algebraic equation that can solved for the series' coefficients. For smooth functions, the convergence of the spectral series is exponential, promoting a high level of accuracy in the solutions.

There is no unique way to define the spectral expansion; in principle, any set of orthogonal polynomials can be used, and most of them have similar convergence properties. The optimal choice ultimately depends  on the domain of integration and the nature of the solution. Fortunately, there exists a significant body of empirical evidence to help optimize the choice of basis polynomials. We refer the reader to Ref.\ \cite{boyd2001chebyshev} for more details.

For the expansion of the function $\delta\! f(p^\mu,\xi)$, the first thing one must consider is the choice of independent variables to represent the momentum $p^\mu$. The symmetries of Eq.\ (\ref{eq:BESimplified}) dictate that $\delta\! f$ should be independent of the azimuthal angle, determined by $p_y/p_x$. Moreover, it greatly simplifies the equations to choose variables that are independent of the position,  Lorentz invariant, that can efficiently represent the fluid's state, and that are defined on a simple domain. A simple choice that satisfies these criteria is the component parallel to the wall $p_\parallel$, and the $z$-component in the frame of the fluid at $\xi=0$, $\mathcal{P}_{\plasma_0}\equiv -p_\mu {\bar u}_{\plasma_0}^\mu$ with ${\bar u}_{\plasma_0}^\mu\equiv {\bar u}_\plasma^\mu|_{\xi=0}$.

For the system of basis functions, we choose the Chebyshev polynomials with an appropriate change of variables to map the infinite domain to $[-1,1]$. For the momenta, we choose the exponential mappings 
\be
\rho_z(\mathcal{P}_{\plasma_0})=\tanh\lp\frac{\mathcal{P}_{\plasma_0}}{2T_0}\rp,\quad \rho_\parallel(p_\parallel)=1-2e^{-p_\parallel/T_0}\,,
\ee
with $T_0\equiv T_{\xi=0}$,\footnote{We use $T_0$ instead of $T(\xi)$ to avoid having  $\xi$-dependence in the momentum mapping. Since $T(\xi)$ varies only slightly across the wall, this does not significantly affect the convergence of the spectral method.} which ensure that $\delta\! f$ decays asymptotically like $e^{-p/T_0}$ for $p\rightarrow\infty$. Since the decay length in the $\xi$ direction is \textit{a priori} unknown, we prefer to use an algebraic mapping for this variable, as it offers more flexibility than the exponential one:
\be
\chi(\xi) = \frac{\xi}{\sqrt{\xi^2+L_\xi^2}}\,,
\ee
where $L_\xi$ is a constant that should be similar in magnitude to the decay length of $\delta\! f$ in the $\xi$ direction.

Expressed in this set of variables, and omitting the species index, the spectral expansion of $\delta\! f$ is 
\be\label{eq:spectralExpansion}
\delta\! f(\chi,\rho_z,\rho_\parallel) = \sum_{i=2}^M\sum_{j=2}^N\sum_{k=1}^{N-1} a_{ijk}\,\bar{T}_i(\chi)\,\bar{T}_j(\rho_z)\,\widetilde{T}_k(\rho_\parallel)\,,
\ee
where $\bar{T}$ and $\widetilde{T}$ are two sets of restricted Chebyshev polynomials defined by
\bea\label{eq:restrictedBasis}
\bar{T}_i(x) &=& \lb T_i(x)-T_0(x)\,,\quad i\ \mathrm{even} \atop T_i(x)-T_1(x)\,,\quad i\ \mathrm{odd} \right. \nn\\
\widetilde{T}_i(x) &=& T_i(x)-T_0(x)\,,
\eea
which are constructed in such a way that $\delta\! f$ automatically satisfies the boundary conditions 
\be
\delta\! f(\xi\rightarrow\pm\infty,\mathbf{p})=\delta\! f(\xi,|\mathbf{p}|\rightarrow\infty) = 0\,.
\ee

The linearized collision integral $\mathcal{C}_\mathrm{lin}[\delta\! f]$ can also be expressed as a spectral series by using Eq.\ (\ref{eq:spectralExpansion}) and the properties of linear operators:
\bea\label{eq:collisionOp}
\mathcal{C}[\delta\! f] &=& \mathcal{C}\lc \sum_{i=2}^M\sum_{j=2}^N\sum_{k=1}^{N-1} a_{ijk}\,\bar{T}_i(\chi)\,\bar{T}_j(\rho_z)\,\widetilde{T}_k(\rho_\parallel) \rc\nn\\
&=& \sum_{i=2}^M\sum_{j=2}^N\sum_{k=1}^{N-1} a_{ijk}\,\bar{T}_i(\chi)\, \mathcal{C}\!\lc \bar{T}_j(\rho_z)\,\widetilde{T}_k(\rho_\parallel) \rc \,.
\eea
The functions $\mathcal{C}\!\lc \bar{T}_j(\rho_z)\,\widetilde{T}_k(\rho_\parallel)\rc$ depend only on the momenta, and not on $\delta\! f$; hence one need compute them only once. 
These functions can be approximated as another spectral series
\be\label{eq:collisionExpansion}
\mathcal{C}\!\lc \bar{T}_j(\rho_z)\,\widetilde{T}_k(\rho_\parallel) \rc \cong \sum_{l=2}^N\sum_{m=1}^{N-1} c_{lm}^{\,jk}\,\bar{T}_l(\rho_z)\,\widetilde{T}_m(\rho_\parallel)\,,
\ee
where the coefficients $c_{lm}^{\,jk}$ can be determined using the method described below. 

The Boltzmann equation (\ref{eq:BESimplified}) can finally be written in terms of the spectral expansions (\ref{eq:spectralExpansion}) and (\ref{eq:collisionOp},\ref{eq:collisionExpansion}) as
{\small
\bea\label{eq:BEspectral}
 &&0=\mathcal{S}(\chi,\rho_z,\rho_\parallel)-\sum_{ijk} a_{ijk}\lb \bar{T}_i(\chi)\sum_{lm} c_{lm}^{\,jk}\,\bar{T}_l(\rho_z)\,\widetilde{T}_m(\rho_\parallel)\right.\nn\\
 &&+\partial_\xi\chi\lc \mathcal{P}_w\partial_\chi-\frac{\gamma_w}{2}\partial_\chi(m^2)(\partial_{p_z}\rho_z)\partial_{\rho_z} \rc\bar{T}_i(\chi)\,\bar{T}_j(\rho_z)\,\widetilde{T}_k(\rho_\parallel)\Bigg\}\,.\nn\\
\eea}

There exist several methods to determine the coefficients $a_{ijk}$. All of them aim at minimizing a residue function that measures the error of Eq.\ (\ref{eq:BEspectral}) (see Ref.\ \cite{boyd2001chebyshev} for more details). The two most common are the {Galerkin} method, which consists of taking moments of the differential equation and setting them to zero, and the {collocation} (or {pseudospectral}) method, which requires the equation to be exactly satisfied on a discrete grid of well-chosen points. One can show that both algorithms approximately minimize the residue
\be\label{eq:residue}
\int_{-1}^1 d\chi d\rho_z d\rho_\parallel\, w(\chi)w(\rho_z)w(\rho_\parallel)\times\left[  \mathrm{Eq.\ (\ref{eq:BEspectral})} \right]^2,
\ee
where $w(x)=1/\sqrt{1-x^2}$ is the weight function under which the Chebyshev polynomials are orthogonal. Both methods have similar convergence properties, but the collocation method does not require carrying out the integral (\ref{eq:residue}); hence we choose collocation.

To minimize the residue (\ref{eq:residue}), one can show that the optimal collocation grid is given by the abscissas of the Gaussian quadrature associated with the Chebyshev polynomials. In the most common version of Gaussian quadrature, these points are the roots of $T_{N+1}$. However, it is more convenient to use instead the Gauss-Lobatto points, which are given by the extrema and endpoints of $T_{N+1}$. The collocation grid is therefore formed by the points $(\chi^{(i)},\rho_z^{(j)},\rho_\parallel^{(k)})$, with\footnote{The collocation grid does not include the boundary points $\chi,\rho_z=\pm 1$ or $\rho_\parallel=1$ because our choice of restricted basis Eq.\ (\ref{eq:restrictedBasis}) automatically satisfies the boundary conditions at these points.}
\bea\label{eq:grid}
\chi^{(i)} &=& \cos\lp\frac{\pi i}{M}\rp,\quad i=1,\cdots,M-1,\nn\\ 
\rho_z^{(j)} &=& \cos\lp\frac{\pi j}{N}\rp,\quad j=1,\cdots,N-1,\nn\\ 
\rho_\parallel^{(k)} &=& \cos\lp\frac{\pi k}{N-1}\rp,\quad k=1,\cdots,N-1.\\
\nn
\eea
Requiring Eq.\ (\ref{eq:BEspectral}) to be satisfied on the grid (\ref{eq:grid}) yields $(M-1)(N-1)^2$ linear algebraic equations that can easily be solved for the $(M-1)(N-1)^2$ unknown coefficients $a_{ijk}$ by doing a single matrix inversion.

The accuracy of this spectral method is expected to increase exponentially with $M$ and $N$. One can generally obtain an error of less than 1\%, which we judge to be satisfactory, with $M\sim20$ and $N\sim10$. This allows for a fast and accurate solution of the Boltzmann equation. 

{It may not be obvious in what ways this spectral method is superior to the standard moment expansion of
the Boltzmann equation, used in previous studies. Seemingly, the latter is closely related to the Galerkin method by a change of weight functions and a trivial basis transformation. Indeed, if performed with exact arithmetic, these two methods show similar convergence properties. The superiority of the Galerkin or collocation methods manifests itself when numerical algorithms are used to do the matrix inversion. For large $N\gtrsim 5$, the high-order terms in the basis functions of the moment expansion become nearly linearly dependent. This results in ill-conditioned matrices that yield large round-off errors when inverted \cite{boyd2001chebyshev}. The Galerkin and collocation methods avoid this problem by using orthogonal polynomials which are, in some way, maximally linearly independent. To prevent the round-off error from becoming prohibitively large when $N$ is large, one should therefore avoid the moment expansion in favor of one of the spectral methods presented in this section. 

There are further reasons to prefer the collocation method over the moment expansion. As previously mentioned, the former does not require integrating the Boltzmann equation, which significantly reduces the numerical overhead. Moreover, several previous studies used additional approximations to simplify the moment equations ({\it e.g.,} neglecting the mass-dependence and the terms proportional to $\partial_\xi(m^2)\delta\! f$). By not having to perform any integration, these approximations become unnecessary and one can easily retain the full Boltzmann equation. This further improves the overall accuracy of the algorithm.}

\begin{figure}[t]
    \centering
    \includegraphics[width=1\linewidth]{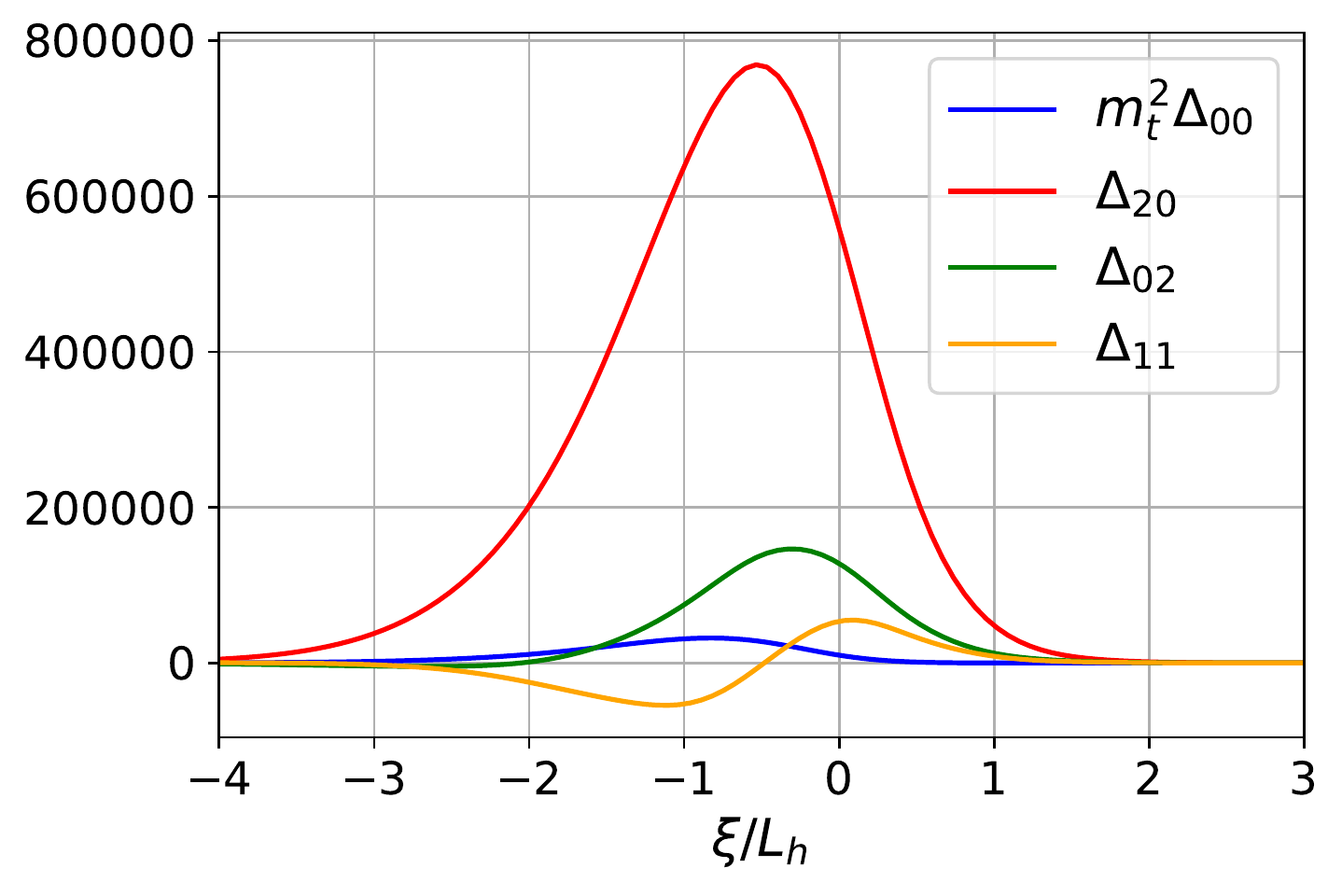}
    \caption{Exemplary solution of the Boltzmann equation for the top quark out-of-equilibrium perturbation $\delta\! f_t$, computed with the spectral method using $N=11$ and $M=22$. We show the moments of $\delta\! f_t$ defined in Eq.\ (\ref{eq:Delta}), which appear in the hydrodynamic equations. The units of the $y$ axis are $\mathrm{GeV}^4$.
    The Lorentz-invariant wall coordinate $\xi$ is defined above Eq.\ (\ref{eq:derivative}).}
    \label{fig:Deltas}
\end{figure}

An example of the results from this procedure is shown in Fig.\ \ref{fig:Deltas}, for the moments of the top quark perturbation $\delta f_t$ as a function of 
$\xi$.  The numbers of basis polynomials $N=11$ and $M=22$ were found to give a high degree of convergence, resulting in an average error of only 0.1\%.  The solution shown corresponds to values of the model parameters, to be described in the following
section: $m_s=103.8\ \mathrm{GeV}$, $\lambda_{hs}=0.72$ and $\lambda_s=1$. The terminal wall velocity for this model is $v_w=0.57$.

\section{Benchmark model: $Z_2$-symmetric singlet scalar extension}
\label{sect:xSM}
To illustrate the methodology discussed earlier, we consider the $Z_2$-symmetric singlet scalar extension of the SM. This model can render the electroweak phase transition strongly first order with a modest input of new physics \cite{McDonald:1993ey,Choi:1993cv,Espinosa:2007qk,Profumo:2007wc,Espinosa:2011ax,Cline:2013gha,Damgaard:2015con,Kurup:2017dzf,Chiang:2018gsn}, which makes it attractive for studying general properties of FOPTs. Moreover, it has been shown that it can generate gravitational waves that could potentially be probed by future detectors \cite{Kakizaki:2015wua,Hashino:2016rvx,Chala:2016ykx,Hashino:2016xoj,Vaskonen:2016yiu,Beniwal:2017eik,Ahriche:2018rao}, and it can provide a successful mechanism for baryogenesis by coupling it to a simple source of $CP$-violation \cite{Cline:2012hg,Cline:2017qpe,Cline:2020jre,Cline:2021iff,Lewicki:2021pgr}.

The singlet scalar extension consists of augmenting the SM by a new scalar field $s$, a singlet under the SM gauge group, and coupling it to the Higgs boson $h$. In its $Z_2$-symmetric version, the tree-level scalar potential of this model takes the general form
\be
V_\mathrm{tree}(h,s) = \frac{\mu_h^2}{2}h^2+\frac{\lambda_h}{4}h^4+\frac{\mu_s^2}{2}s^2+\frac{\lambda_s}{4}s^4+\frac{\lambda_{hs}}{4}h^2s^2\,.
\label{eq:vtree}
\ee
To make quantitative predictions, we add to this potential the one-loop vacuum and thermal corrections, which are described in the appendix of Ref.\ \cite{Friedlander:2020tnq}.

We are interested in the region of parameter space where the phase transition occurs in a two-step process, which first breaks the $s$ field's $Z_2$ symmetry and subsequently that of the Higgs field. Electroweak symmetry breaking occurs at the second step, so we only consider the second phase transition in the following. Electroweak bubbles appear at the nucleation temperature $T_n$, which is always below the critical temperature $T_c$ where the two vacua are degenerate. 
Further details about the EWPT and  bubble nucleation in the singlet scalar extension can be found in Refs.\ \cite{Ham:2004cf,Kozaczuk:2015owa,Kurup:2017dzf,Chiang:2018gsn,Friedlander:2020tnq,Cline:2021iff,Lewicki:2021pgr,Guo:2021qcq}.

\subsection{Solution of fluid equations for SM-like plasma}
To determine the wall velocity $v_w$, one needs to solve simultaneously a set of equations consisting of the scalar fields' EOMs (\ref{eq:EOM}), the conservation of the EMT (\ref{eq:EMTsimplified}) and the spectral Boltzmann equation (\ref{eq:BEspectral}). In practice, the out-of-equilibrium perturbations contribute only slightly to the EOMs and the EMT. This allows one to solve the whole system iteratively, by first solving the EOMs and EMT with $\delta\! f=0$, then using the resulting fields, temperature and velocity profiles to solve the Boltzmann equation for an updated $\delta\! f$. We iterate this process with the new $\delta\! f$ until convergence is achieved. This algorithm allows one to separate the problem into solutions of two subsystems: the spectral Boltzmann equation which can be solved with the method described in Section \ref{sect:spectral}, and the EOMs+EMT system. We show how to solve the latter in this subsection.

To simplify the analysis, we only consider the out-of-equilibrium contribution from the top quark, which is the dominant one.  In a more complete treatment,  the other  massive degrees of freedom including the $W$ and $Z$ bosons and the scalar fields should also be taken into account. 
We defer such improvements to future investigation.
With this simplification, the scalar fields' EOMs in the wall frame become
\bea\label{eq:EOMtop}
E_h&=&-\partial_z^2 h+\frac{\partial V_T(h,s;T)}{\partial h}+N_t\frac{\partial (m_t^2)}{\partial h}\frac{\Delta_{00}^t}{2}=0\,,\nn\\
E_s&=&-\partial_z^2 s+\frac{\partial V_T(h,s;T)}{\partial s}=0\,,
\eea
where $V_T$ is the effective potential including the one-loop vacuum and thermal corrections, and $N_t=12$ is the top quark's number of degrees of freedom.

\begin{figure*}[t!]
\centerline{\includegraphics[width=0.45\linewidth]{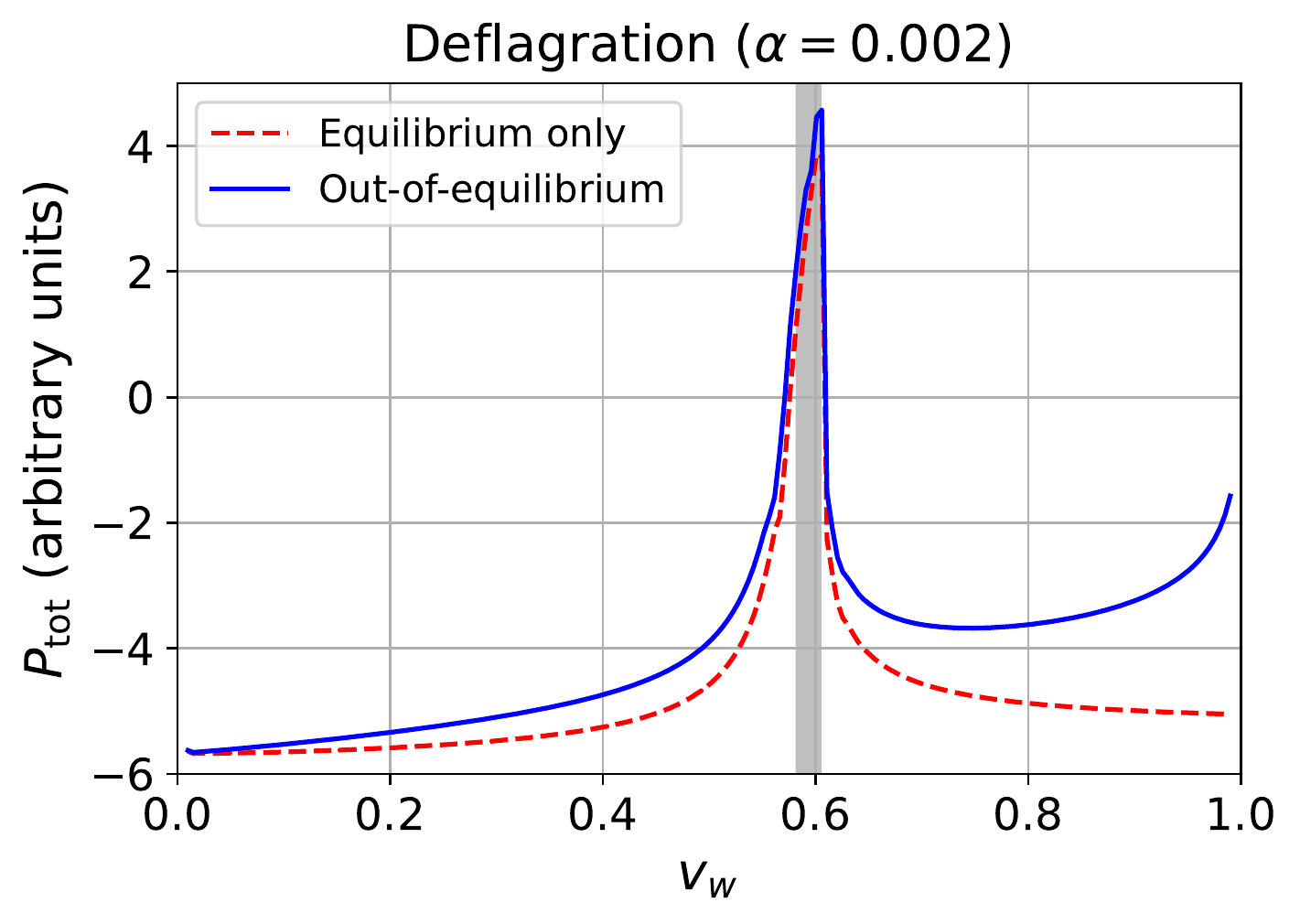}\hspace{0.05\linewidth}\includegraphics[width=0.45\linewidth]{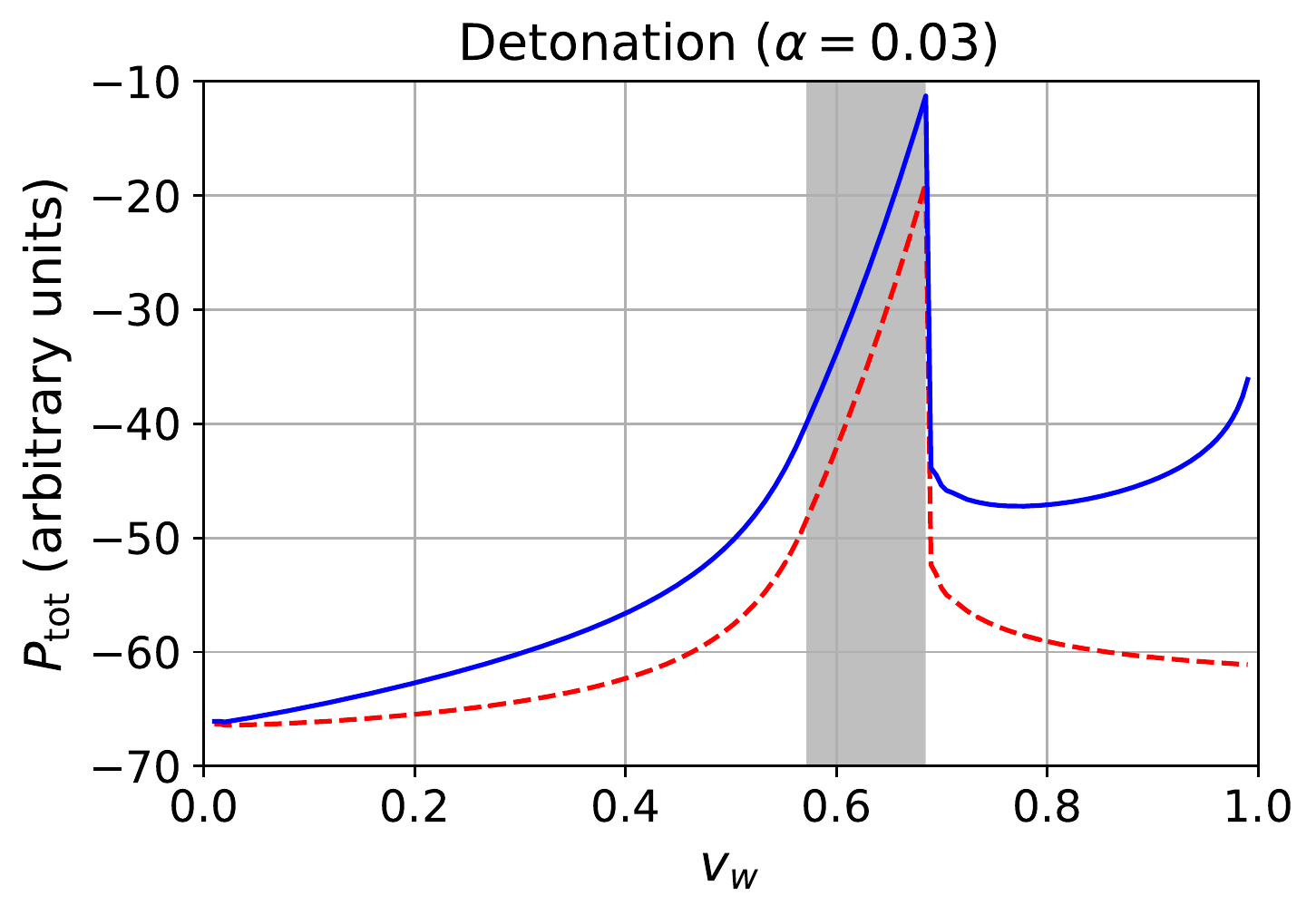}}
\centerline{(a)\hspace{0.48\linewidth}(b)}
 \caption{Total pressure as a function of the wall velocity for a deflagration wall (a) and a  detonation wall (b). The dashed (red) line only contains the contribution from the equilibrium distribution function, while the solid (blue) line also includes the top quark's out-of-equilibrium contribution. The shaded region corresponds to hybrid walls. It is bounded on the right by the Jouguet velocity, $v_J$.} 
 \label{fig:pressure}
\end{figure*}

A simple approximate solution to these equations is the $\tanh$ {ansatz}
\bea\label{eq:ansatz}
h(z)&=&\frac{h_0}{2}[1-\tanh(z/L_h)]\,,\nn\\
s(z)&=&\frac{s_0}{2}[1+\tanh(z/L_s+\delta_s)]\,,
\eea
where $h_0$ and $s_0$ are the scalar field VEVs in the true and false vacua, respectively, given appropriate choices for the coupling constants in the potential (\ref{eq:vtree}).  They satisfy
\bea
\frac{\partial V}{\partial h}(h_0,0;T_-)&=&\frac{\partial V}{\partial h}(0,s_0;T_+)=0\,,\nn\\
\frac{\partial V}{\partial s}(h_0,0;T_-)&=&\frac{\partial V}{\partial s}(0,s_0;T_+)=0\,.\nn\\
\nn
\eea
The VEVs depend on the plasma's asymptotic temperatures
 $T_\pm$ far from the wall, which are themselves a function of the wall velocity $v_w$. 

Once a tentative field profile has been proposed, one can determine the temperature profile by solving Eq.\ (\ref{eq:EMTsimplified}). Since no derivative of $T$ is involved, it forms a set of uncoupled algebraic equations for each value of $z$. Therefore,  it can be solved with a standard root-finding algorithm, like a Newton or bracketing method.\footnote{For some tentative field profiles and wall velocities (mainly for hybrid walls), Eq.\ (\ref{eq:EMTsimplified}) does not have any solution. In this situation, we choose the value of $T$ that minimizes the error. Of course, the exact solution of the EOMs (\ref{eq:EOMtop}) must allow a solution for Eq.\ (\ref{eq:EMTsimplified}).}

To get an accurate estimate of the solution, it is essential to allow the two scalar fields to have different wall thicknesses $L_h$ and $L_s$, and even more so to allow an offset $\delta_s$ between the two wall positions. One could also use a more general {ansatz} than Eq.\ (\ref{eq:ansatz}) with more free parameters. However, we find that the main features of the fields' profiles are well approximated by (\ref{eq:ansatz}), and it is therefore sufficient for obtaining a good enough initial estimate of the solution.

The general strategy for determining the optimal values of $v_w$, $L_h$, $L_s$ and $\delta_s$ is to take moments of
$E_h$ and $E_s$ in Eq.\ (\ref{eq:EOMtop}) and algebraically solve 
for the vanishing of the moments. 
The choice of the moments is to some extent arbitrary,
as long as they are sensitive to independent linear combinations of the unknown parameters.
A convenient choice is \cite{Konstandin:2014zta}
\bea\label{eq:EOMmoments}
P_h(v_w,L_h,L_s,\delta_s) &=& -\int \!dz\, E_h h'=0\,,\nn\\
G_h(v_w,L_h,L_s,\delta_s) &=& \int \!dz\, E_h(2h/h_0-1)h'=0\,,\quad
\eea
and similarly for $P_s$ and $G_s$. These moments have intuitive physical interpretations that naturally distinguish them as good predictors of the wall speed and thickness, respectively. $P_i$ is a measure of the net pressure on the wall, so that $P_i=0$ can be interpreted as the requirement that a stationary wall should have a vanishing total pressure; nonvanishing $P_i$ would cause it to accelerate. It can also create an offset between the two walls if $P_h\neq P_s$, which can be used to determine $\delta_s$. On the other hand, $G_i$ measures the pressure gradient in the wall. If nonvanishing, it would lead to compression or stretching of the wall, causing $L_i$ to change.

The system of moment equations (\ref{eq:EOMmoments}) can be solved with a Newton algorithm. If needed, the resulting approximate solution can then be improved with a few relaxation steps. Generally, we find that it only changes the wall velocity by a few percent, so the tanh {ansatz} is sufficient for most applications.

\subsection{Classification of solutions}

To better understand how the wall velocity of first-order phase transition bubbles is determined by the departure from equilibrium, it is convenient to define the total pressure
\be
P_\mathrm{tot}(v_w)=P_h(v_w,L_h^*,L_s^*,\delta_s^*)+P_s(v_w,L_h^*,L_s^*,\delta_s^*),
\ee
where $L_h^*$, $L_s^*$ and $\delta_s^*$ solve the equations $P_h-P_s=G_h=G_s=0$. This quantity measures the total pressure on the bubble wall as a function of the wall velocity, when the dependence on the wall shape is eliminated in favor of $v_w$.

The total pressure for two different models is shown in Fig.\ \ref{fig:pressure}, illustrating deflagration and detonation transitions, respectively.
A striking nontrivial feature in both curves is the large pressure peak maximized at the Jouguet velocity $v_J$, which is the smallest wall velocity that can yield a detonation solution. This pressure barrier was  predicted in Ref.\ \cite{Konstandin:2010dm} and observed in Ref.\ \cite{Cline:2021iff}, and is confirmed by the present analysis. It plays an essential role, being a general feature of first-order phase transitions, and naturally dividing the bubbles into two qualitatively distinct categories: deflagration (which also includes hybrid walls) and ultrarelativistic detonation solutions.\footnote{For more formal definitions of deflagration, hybrid and detonation solutions, see Appendix \ref{app:BC}.}

A phase transition corresponds to a deflagration solution if it is too weak to overcome the pressure barrier at $v_J$. This is the case for the model shown in Fig.\ \ref{fig:pressure}(a), which solves $P_\mathrm{tot}=0$ at approximately $v_w=0.57$. It is no coincidence that the terminal wall velocity in this example ends up being so close to the sound speed in the plasma. The pressure peak is a consequence of hydrodynamic effects that heat the plasma, increasing the pressure on the wall. These effects become especially important for hybrid walls, when the shock wave in front of the wall becomes thin. This causes the pressure to start to rise rapidly only around $v_w\sim c_s\sim 0.58$, when the wall becomes a hybrid solution (a supersonic deflagration which has both a shock and rarefaction wave). Since the acceleration of most deflagration solutions is stopped by the pressure peak, it implies that a large fraction of the walls in that category satisfies $c_s \lesssim v_w\leq v_J$.

For stronger phase transitions, the pressure peak is not high enough to impede the wall's acceleration, since $P_{\rm tot}$ remains negative, leading to detonation solutions. This is the case for the model shown in Fig.\ \ref{fig:pressure}(b). Since the pressure at $v_J$ is higher than in the range $v_J<v_w\lesssim1$ and yet insufficient to decelerate the wall, it will continue accelerating until it is stopped by the friction coming from out-of-equilibrium effects. These  become important only at ultrarelativistic velocities; in general, detonation solutions always satisfy $\gamma_w\gtrsim 10$. The asymptotic behavior of this ultrarelativistic friction force is still being debated, as Refs.\ \cite{Bodeker:2017cim,Hoche:2020ysm}  found different scaling relations $P_\mathrm{tot}(v_w\rightarrow1)\sim \gamma_w,\gamma_w^2$, respectively. Nevertheless, they both agree that it will eventually become high enough to stop the wall with $\gamma_w\gg1$.

One can also appreciate the role played by the deviation from equilibrium in Fig.\ \ref{fig:pressure}, as it shows the pressure computed with or without the out-of-equilibrium contribution (the term proportional to $\Delta_{00}$ in Eq.\ (\ref{eq:EOMtop})). Strikingly, the out-of-equilibrium pressure is much smaller than the equilibrium one.\footnote{\label{fn7}However, one should keep in mind that we only include the out-of-equilibrium contribution of the top quark, which slightly underestimates the actual total pressure.} The latter can accurately reproduce the most important feature of the pressure curve, namely the pressure peak at $v_J$. Therefore, it could be a good approximation to neglect the deviations from equilibrium altogether, considerably simplifying the fluid equations since no Boltzmann equation would have to be solved.

However, there are some situations where the out-of-equilibrium pressure can be quantitatively important. For slow walls ($v_w\lesssim 0.5$), the error in the wall velocity $\Delta v_w$ caused by neglecting this contribution can become large, even if the 
pressure difference is small. One can show that $\Delta v_w\sim \lp dP_\mathrm{tot}/d v_w\rp^{-1}$, which increases at small velocities. But as previously argued, most deflagration solutions are stopped by the pressure peak, at which point $\lp dP_\mathrm{tot}/d v_w\rp^{-1}$
is small; thus, $\Delta v_w$ is expected to be unimportant for typical deflagrations. 

One can also see from Fig.\ \ref{fig:pressure} that the approximation of ignoring the out-of-equilibrium pressure contribution becomes poor when $v_w>v_J$. But ultimately this has a small effect, since all detonation walls become ultrarelativistic, irrespective of their pressure profile. Possibly the most important distinction between the equilibrium and total pressure is the magnitude of the peak pressure, which determines the type of solution: deflagration for $P_\mathrm{tot}(v_J) > 0$ and detonation otherwise. Even if the out-of-equilibrium contribution is small, it is possible that neglecting it erroneously transforms a deflagration into a detonation, resulting in a large $\Delta v_w$. We will later estimate how likely this is to happen and propose a simple fix to approximate the out-of-equilibrium pressure without solving the Boltzmann equation.

\begin{figure}[t]
    \centering
    \includegraphics[width=1\linewidth]{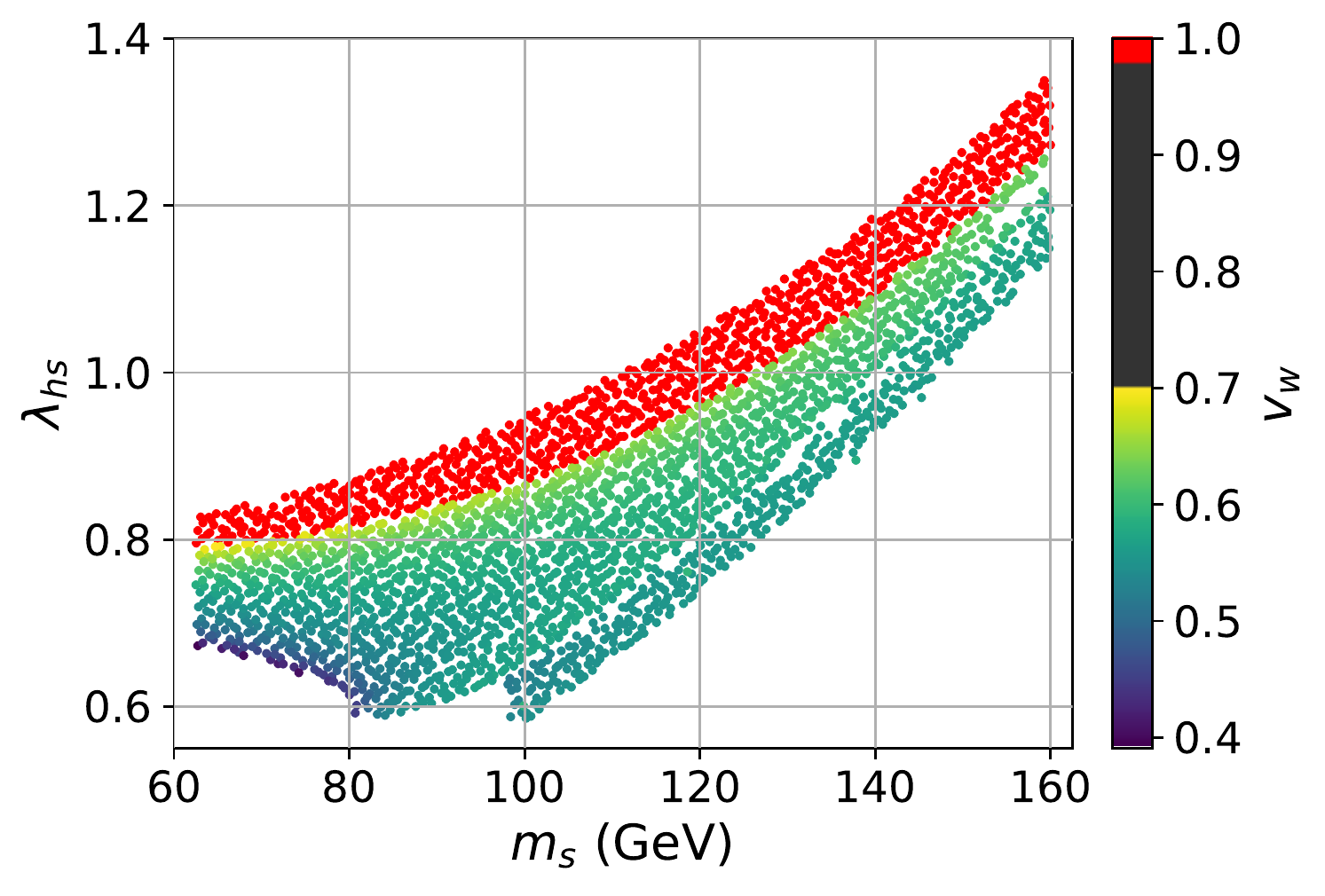}
    \caption{Scan of the parameter space with $\lambda_s=1$. Red points are ultrarelativistic detonation solutions.}
    \label{fig:scan}
\end{figure}

\subsection{Scan results}

We now turn to a more specific discussion about the singlet scalar extension. To study the consequences of the methodology discussed in this paper for the wall velocity and shape, we performed a random scan over the region in parameter space constrained by $\lambda_s=1$ and $m_s\in [62.5,160]\ \mathrm{GeV}$, where $m_s=\mu_s^2-\lambda_{hs}\mu_h^2/(2\lambda_h)$ is the physical mass of the $s$ particle at $T=0$, and its lower bound is chosen to avoid collider constraints from Higgs boson decays $h\to ss$.   To ensure a more uniform exploration of the parameter space, we sampled the points from a Sobol sequence, which prevents two points from being arbitrarily close to one another. The result of this scan is shown in Fig.\ \ref{fig:scan}, where $v_w$ is indicated in the $\lambda_{hs}$-$m_s$ plane for models giving a first order phase transition.

Of the 2860 points sampled, roughly 70\% are deflagration walls. Although this ratio is model-dependent, it suggests that both types of solution should be relatively common for generic models. The qualitative analysis of the last subsection is validated: we find that all the detonation solutions have $\gamma_w>10$ (hence ultrarelativistic), 
and a large fraction of the deflagration walls have a terminal velocity close to the speed of sound, the slowest being $v_w\approx0.4$. More precisely, 83\% of the deflagration walls have a velocity greater than 0.55, and 97\% are greater than 0.5.

\begin{figure}[t]
    \centering
    \includegraphics[width=1\linewidth]{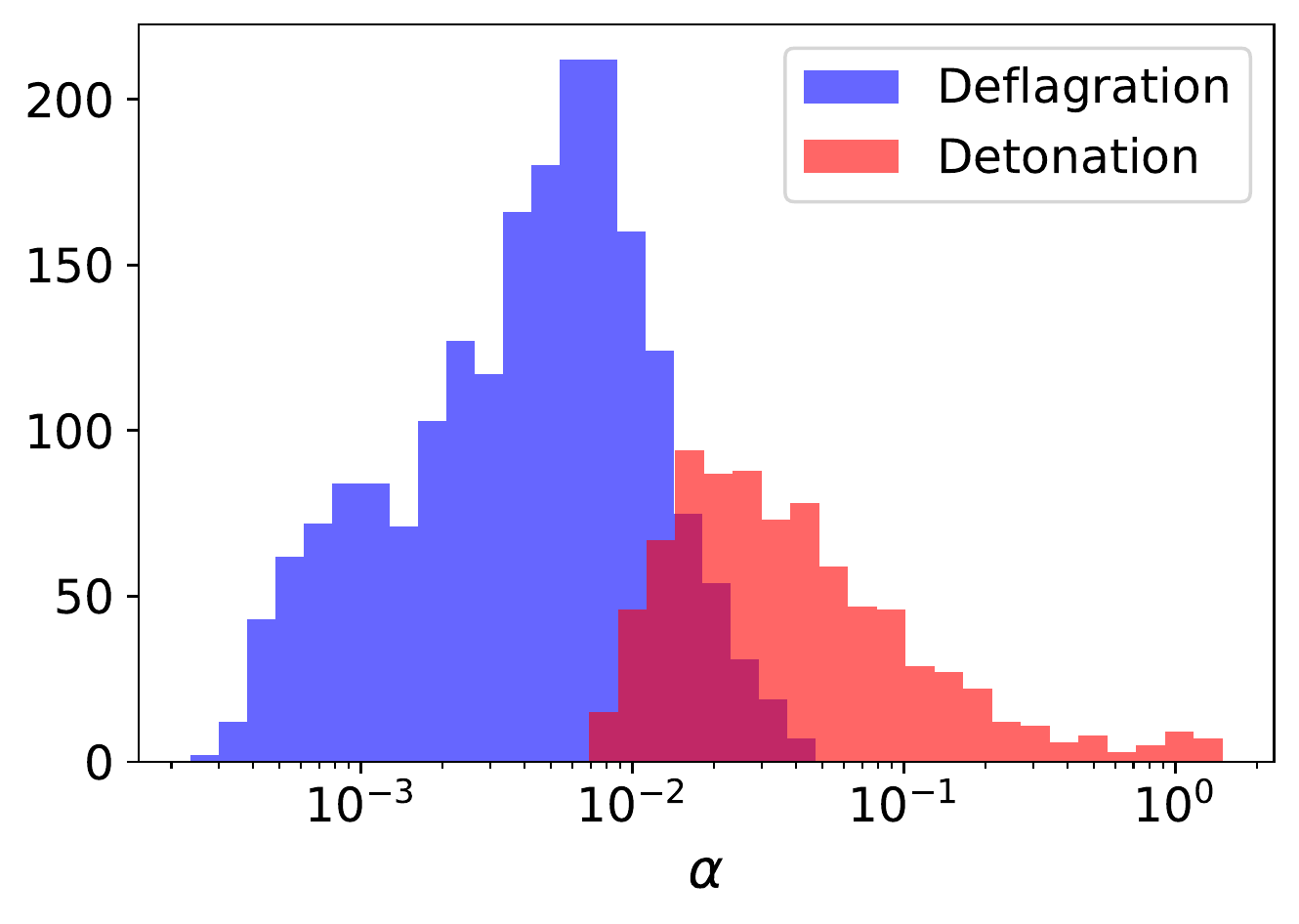}
    \caption{Histogram of $\alpha$ (latent heat divided by radiation energy density) for deflagration (blue) and detonation (orange) solutions.  Overlap region is shown in darker red.}
    \label{fig:alpha}
\end{figure}

The classification of walls is strongly correlated with the strength of the phase transitions, which we quantify by $\alpha$, the ratio of released vacuum energy density to the radiation energy density. A histogram of $\alpha$ is shown in Fig.\ \ref{fig:alpha}. It clearly shows that the phase transition of detonation solutions is, in general, stronger than for deflagration walls. There is some overlap between the two groups, but as a rough rule of thumb, one can say that deflagrations satisfy $\alpha_\mathrm{def}\lesssim 10^{-2}$ and detonations $\alpha_\mathrm{det}\gtrsim10^{-2}$.

\begin{figure}[t]
    \centering
    \includegraphics[width=1\linewidth]{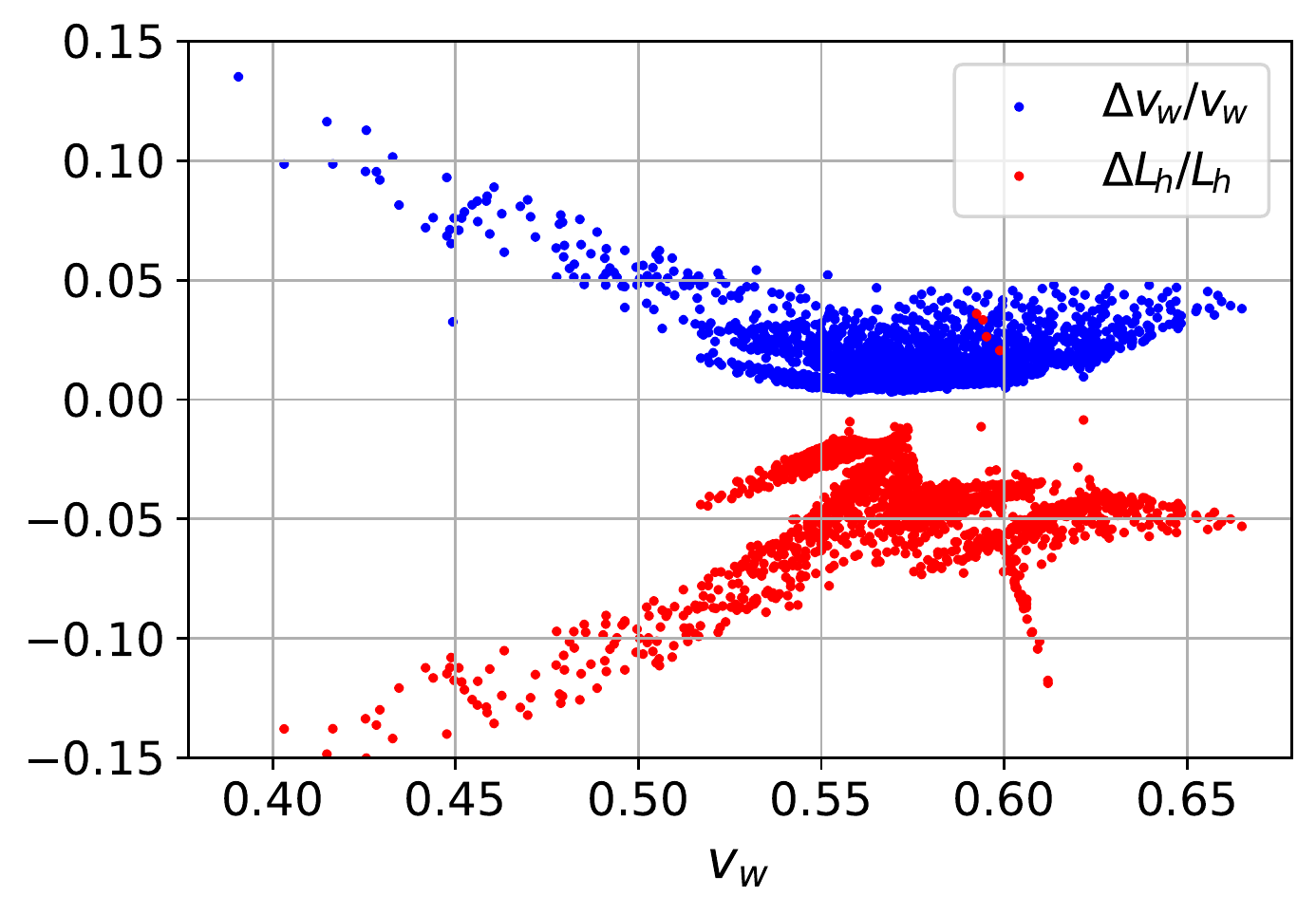}
    \caption{Scatter plot of relative errors of the wall velocity (blue points) and thickness (red points) due to neglecting the out-of-equilibrium pressure contribution, as a function of $v_w$.}
    \label{fig:error}
\end{figure}

We quantify  the effect of neglecting the out-of-equilibrium pressure in Fig.\ \ref{fig:error}, which shows the relative errors induced for the wall velocity and thickness. The absolute error is defined as $\Delta v_w=v_w^\mathrm{eq}-v_w^\mathrm{total}$, and similarly for $L_h$; thus Fig.\ \ref{fig:error} shows that the out-of-equilibrium contributions slow down the wall while making it thicker. As expected, the error is small for hybrid walls and becomes larger as $v_w$ decreases. Even in the extreme cases, the relative error never exceeds 15\% for $v_w$ and 20\% for $L_h$, and the mean errors are 2\% and 5\%, respectively. 

The out-of-equilibrium pressure also has an impact on the classification of the solutions, as 9\% of the deflagration walls would be incorrectly identified as detonations if one neglected this contribution. This last consequence could be more problematic since it has a significant qualitative impact on the behavior of the misidentified solution.
The errors presented here somewhat underestimate the true error that would be obtained by considering the full out-of-equilibrium contributions beyond those of the top quark.$^{\ref{fn7}}$

\section{Approximation schemes}
\label{sect:approx}

For some applications, a qualitative understanding of the bubble wall's behavior may be sufficient; for example one might be content to know whether the wall speed is subsonic or ultrarelativistic, without needing its precise value.
With that in mind, in this section we propose several approximation schemes with increasing levels of complexity and accuracy. Each method can be separated into two parts: first classifying the wall as a deflagration or detonation solution, and second computing the wall velocity\footnote{For detonations, the wall velocity is always $v_w\approx 1$ so this second part becomes trivial. For estimations of the terminal Lorentz factor $\gamma_w$ for detonation solutions, see Refs.\ \cite{Bodeker:2017cim,Hoche:2020ysm}}.\\

\subsection{Fixed wall velocity}
The wall velocity is often needed to study the cosmological signatures of first-order phase transitions ({\it e.g.,} baryogenesis and gravitational waves). To simplify the analysis, several past studies have assumed a fixed value for the wall velocity instead of computing it from the fluid equations. Unfortunately, the chosen value was often motivated by the strength of the cosmological signature studied instead of the likelihood of having sucha  velocity, simply because this likelihood was unknown. 

For example, baryogenesis studies frequently adopt a small wall velocity ($v_w\lesssim0.1$) to maximize the resulting baryon asymmetry. As we have seen in Section \ref{sect:xSM}, it is actually very unlikely to have $v_w<0.5$, and the smallest velocity found out of 2860 models is $v_w=0.4$. Instead of this arbitrary (and inaccurate) assumption, we propose here to use the most likely wall velocity, which is $v_w=c_s=1/\sqrt{3}$ for deflagrations and $v_w=1$ for detonations. It is harder to suggest a fixed wall thickness since this typically depends rather sensitively on the specific model being studied. Generally, values between $5/T_n$ and $20/T_n$ are realistic. One can classify the solutions using the $\alpha$ parameter, with phase transitions satisfying $\alpha<10^{-2}$ corresponding to deflagrations and $\alpha>10^{-2}$ as detonations.

Using these approximations, we correctly classify 86\% of the models sampled in the last section. Moreover, the mean error in the wall velocity of deflagration walls is only 5\%. Of course, the main advantage of this approximation is that it requires almost no calculation. Yet, it is able to reproduce the correct qualitative behavior most of the time. For higher accuracy, or an estimation of the wall thickness, we recommend one of the subsequent approximations.\\

\subsection{Local thermal equilibrium}
The next step towards a better approximation of the fluid equations is to assume the plasma to be in local thermal equilibrium (LTE). This strategy has already been studied in Refs.\ \cite{BarrosoMancha:2020fay,Balaji:2020yrx,Ai:2021kak}, and we showed in Section \ref{sect:xSM} that it correctly reproduces the qualitative features of the pressure curve. To implement it, one needs to solve Eqs.\ (\ref{eq:EOM},\ref{eq:EMTsimplified}) while neglecting all the out-of-equilibrium terms involving the $\Delta_{mn}$ functions. The type of solution can be classified by computing the sign of $P_\mathrm{tot}(v_J)$.

Using a tanh {ansatz} to solve the equilibrium fluid equations, one can in this way correctly classify 94\% of the solutions. The mean error on the deflagration wall velocity is 2\%, and the wall thickness estimate has an accuracy of 13\%.
This level of accuracy represents a net improvement compared to the previous approximation, in addition to providing an estimation of the wall shape. However, it requires computing $v_J$, $v_\pm$, $T_\pm$ and solving a set of fluid equations, which requires significant additional computational effort.\\

\subsection{Numerical fit of out-of-equilibrium pressure}
It is possible to substantially improve the last approximation by estimating the out-of-equilibrium pressure from a numerical fit. Since the Boltzmann equation does not depend on the potential $V_T$, these out-of-equilibrium contributions are, to some extent, model-independent. 

\begin{figure}[t]
    \centering
    \includegraphics[width=1\linewidth]{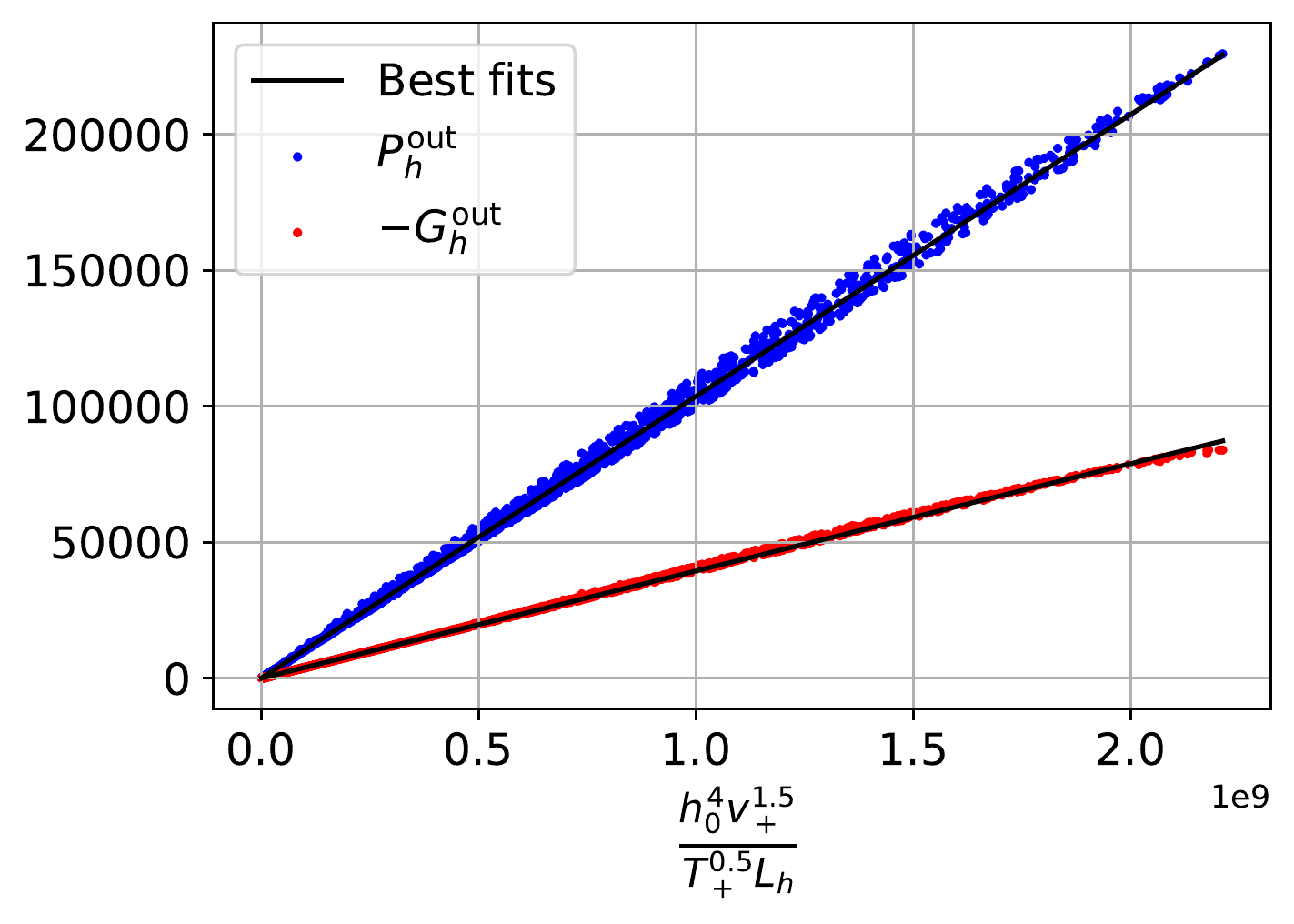}
    \caption{Out-of-equilibrium pressure (blue points) and pressure gradient  (red points) of the deflagration solutions obtained in the scan of Section \ref{sect:xSM}, and the corresponding best fits Eq.\ (\ref{eq:fits}) (lines). The units of the $y$ axis are $\mathrm{GeV}^4$.}
    \label{fig:fits}
\end{figure}

An excellent representation of out-of-equilibrium contributions to the moments of Eq.\ (\ref{eq:EOMmoments}) can be obtained using a power-law proportional to $h_0^4$, $v_+^{1.5}$, $T_+^{-0.5}$ and $L_h^{-1}$ (Recall that $v_+$, $T_+$ are the fluid velocity and temperature at $z\to\infty$.) The best fits of the out-of-equilibrium pressure and pressure gradient are
\bea\label{eq:fits}
P_h^\mathrm{out}&\equiv& -\frac{N_t}{2}\int \!dz\, h'\frac{\partial (m_t^2)}{\partial h}\Delta_{00}^t\nn\\
&\approx& (1.04\times10^{-4}\ \mathrm{GeV}^{-0.5})
\frac{N_t\, y_t^4\,h_0^4\,v_+^{1.5}}{T_+^{0.5}L_h}\,,\nn\\
G_h^\mathrm{out}&\equiv& \frac{N_t}{2}\int \!dz\, (2h/h_0-1)h'\frac{\partial (m_t^2)}{\partial h}\Delta_{00}^t\nn\\
&\approx& (-3.95\times10^{-5}\ \mathrm{GeV}^{-0.5})\frac{N_t\, y_t^4\,h_0^4\,v_+^{1.5}}{T_+^{0.5}L_h}\,,
\eea
where $y_t$ is the top Yukawa coupling. We calibrated these fits on the deflagration walls found in the scan of Section \ref{sect:xSM}, so they should not be trusted for ultrarelativistic detonation solutions. They both have a coefficient of determination of $R^2=0.999$, and the fits are shown in Fig.\ \ref{fig:fits}.

One can substitute the fits (\ref{eq:fits}) in the moment equations (\ref{eq:EOMmoments}) to obtain an improved set of fluid equations that include the estimate of the out-of-equilibrium contributions\footnote{Since this approximation is obtained from the moment equations (\ref{eq:EOMmoments}), it requires a tanh {ansatz} for the fields' profiles}. Using this approximation, we can correctly classify 98\% of the walls and the mean errors on $v_w$ and $L_h$ are respectively 0.4\% and 9.8\%.

At this stage, the numerical fits (\ref{eq:fits}) are only valid for the singlet scalar extension presented in Section \ref{sect:xSM}. It is possible to generalize them by identifying the origin of the $h_0^4/L_h$ factor, which is model-dependent. First, the $dz\, h'\partial(m^2)/\partial h$ factor in the definitions of $P_h^\out$ and $G_h^\out$ scales like $h_0^2$. Second, the magnitude of deviation from equilibrium is proportional to the amplitude of the Boltzmann equation's source term ${\cal S}_i$, Eq.\ (\ref{eq:source}). The dominant contribution to ${\cal S}_i$ is the term proportional to $\partial_\xi (m_t^2)$, which scales like $h_0^2/L_h$.

In a more general treatment, one should include the pressure from all the massive degrees of freedom. Furthermore, the masses can depend on the VEVs of several scalar fields. To be as general as possible, we keep the masses unspecified and replace the $h_0^4/L_h$ factor in Eq.\ (\ref{eq:fits}) by the following scaling relations
\bea\label{eq:beta}
\Delta_{00}^j&\sim& \partial_\xi(m_j^2)\sim\sum_k \frac{\beta_{jk}}{L_k},\nn\\
\beta_{ij}&\equiv& \Delta\!\!\lbr\phi_j\frac{\partial (m_i^2)}{\partial\phi_j}\rbr,
\eea
and
\bea\label{eq:fitsGeneral}
P_i^\out &=& -\sum_j\frac{N_j}{2}\int\! dz\, \phi_i'\frac{\partial(m_j^2)}{\partial\phi_i}\Delta_{00}^j\nn\\
&\sim& \sum_j N_j\, \beta_{ji}\sum_k \frac{\beta_{jk}}{L_k},\qquad\\ \nn
\eea
where the $\Delta$ operator denotes the variation of its argument across the wall, and $L_k$ is the thickness of the $\phi_k$ wall profile. The factors $\beta_{ij}$ can be interpreted as the variation of $m_i^2$ during the phase transition due to the variation of $\phi_j$'s VEV.

The $v_+^{1.5}/T_+^{0.5}$ factor appearing in the fits does not depend on the masses, but rather on the structure of the Boltzmann equation. Therefore, we expect it to be model-independent. This is a reasonable assumption for SM-like plasmas since they all share similar collision operators. 

Demanding that Eq.\ (\ref{eq:fitsGeneral}) reduces to Eq.\ (\ref{eq:fits}) in the case where only the contribution from the top quark is included, the fits for the pressure and pressure gradient become
\bea\label{eq:fitsFinal}
P_i^\out&\approx& \frac{1.04\times10^{-4}}{\mathrm{GeV}^{0.5}}\,\frac{v_+^{1.5}}{T_+^{0.5}}\sum_{j,k}\frac{N_j}{L_k}\beta_{ji}\beta_{jk}\,,\nn\\
G_i^\out&\approx& \frac{-3.95\times10^{-5}}{\mathrm{GeV}^{0.5}}\,\frac{v_+^{1.5}}{T_+^{0.5}}\sum_{j,k}\frac{N_j}{L_k}\beta_{ji}\beta_{jk}\,,
\eea
where $j$ is summed over all the massive species and $k$ over all the scalar fields that have VEVs in the wall.

As an exemplary application, we use the fits (\ref{eq:fitsFinal}) to estimate the out-of-equilibrium contributions from the $h$ and $s$ scalar fields, and the $W$ and $Z$ bosons, which were previously neglected.

The contribution from the gauge bosons is the easiest to estimate since their mass is not affected by the singlet scalar. Their out-of-equilibrium pressure can simply be related to the top's by
\be
\frac{P_{h,W}^\out+P_{h,Z}^\out}{P_{h,t}^\out}\approx \frac{N_W m_W^4+N_Z m_Z^4}{N_t m_t^4}\approx 0.04,
\ee
where $P_{i,j}^\out$ is the out-of-equilibrium pressure of the $j$ species on the $\phi_i$ wall. This shows that it is a reasonable approximation to neglect the gauge bosons' contribution.

The scalar fields' masses depend $h$ and $s$, so they contribute to both $P_h^\out$ and $P_s^\out$. Before estimating these pressures, it is convenient to compute the relevant $\beta_{ij}$ factors from Eqs.\ (\ref{eq:vtree}) and (\ref{eq:beta}):
\bea
\beta_{hh}&=&6\lambda_h h_0^2,\quad\beta_{hs}=\lambda_{hs}s_0^2\,.\nn\\
\beta_{ss}&=&6\lambda_s s_0^2,\quad\,\beta_{sh}=\lambda_{hs}h_0^2\,.
\eea
The fractional contribution from the scalar particles to the pressure, relative to that of the top quark,  can then be estimated as
\bea
\frac{L_h}{N_t(y_t h_0)^4}\lc \frac{1}{L_h}(\beta_{hh}^2+\beta_{sh}^2)+\frac{1}{L_s}(\beta_{hh}\beta_{hs}+\beta_{sh}\beta_{ss}) \rc&\sim& 0.3\nn\\
\frac{L_h}{N_t(y_t h_0)^4}\lc \frac{1}{L_s}(\beta_{ss}^2+\beta_{hs}^2)+\frac{1}{L_h}(\beta_{hh}\beta_{hs}+\beta_{sh}\beta_{ss})\rc&\sim& 0.5\nn\\
\eea
for the $h$ and $s$ walls, respectively. The numbers on the right-hand-sides are averages over all the deflagrations found in Sec.\ \ref{sect:xSM}. Due to the large value of $\lambda_s=1$ used for the scan, the magnitude of this out-of-equilibrium pressure is significantly larger. Incidentally, nearly all the contribution comes from the $s$ field's departure from equilibrium. Therefore, it is still a good approximation to assume the Higgs to be in LTE (because $\lambda_h\sim0.1$ is small), but it seems harder to justify it for the singlet scalar, as long as $\lambda_s$ is large.

We finally study the effect of these additional sources of pressure on the wall velocity. Despite the large contribution from the $s$ field, the average $v_w$ for deflagration solutions is only decreased by 0.5\%, compared to the case where only the top's pressure is included. This is not surprising since we already found that $v_w$ is mainly determined by the equilibrium contribution. However, the additional pressure increases the number of deflagration solutions by 7\%, which could have an important effect on the predicted terminal velocity if a high accuracy is needed.

\subsection{Discussion}
Table \ref{table:accuracy} shows a summary of the accuracy of each approximation scheme. For applications requiring only a qualitative understanding, we recommend the {fixed wall velocity} method, as it requires practically no calculations, while correctly describing the bubble's qualitative dynamics. For higher accuracy, we recommend  solving the fluid equations with the numerical fits (\ref{eq:fitsFinal}) to estimate the out-of-equilibrium terms, since it is much more precise than the {local thermal equilibrium} (LTE) method, without increasing the complexity of the equations that must be solved.  However, these fits should only be trusted for a SM-like plasma; the LTE method should be used for a plasma whose constituents differ signficantly from the SM.

Finally, for the highest level of accuracy, one should compute the deviation from equilibrium by solving a set of Boltzmann equations for each massive species in the plasma. This can be done efficiently with the spectral method described in Section \ref{sect:spectral}.

It is worthwhile to keep in mind that the Boltzmann equation is not an exact predictor of the distribution functions, and is subject to various sources of uncertainty.  It is an approximation to the Kadanoff-Baym equations, which is obtained by performing a gradient expansion \cite{Konstandin:2013caa}, that is  only trustworthy for $L_i\gtrsim 2/T_+$ \cite{Jukkala:2019slc}. Moreover, the collision integrals are computed to leading log accuracy, which has large uncertainties \cite{Moore:1995si,Kozaczuk:2015owa,Laurent:2020gpg,Wang:2020zlf}. This can induce significant errors in the out-of-equilibrium contributions, since they are typically inversely proportional to the collision terms. 

\begin{table}
\centering
\begin{tabular}{ | c | c | c | c | } 
 \hline
 Approximation & Correct classification & $\Delta v_w/v_w$ & $\Delta L_h/L_h$\\ 
 \hline
 Fixed wall & \multirow{2}{*}{86} & \multirow{2}{*}{4.7} & \multirow{2}{*}{-}\\ 
 velocity & & &\\
 \hline
 Local thermal & \multirow{2}{*}{94 (94)} & \multirow{2}{*}{1.6 (2.2)} & \multirow{2}{*}{4.7 (13)} \\ 
 equilibrium & & &\\
 \hline
 Numerical fits & (98) & (0.4) & (9.8) \\ 
 \hline
  Full Boltzmann & \multirow{2}{*}{100 (98)} & \multirow{2}{*}{0 (0.5)} & \multirow{2}{*}{0 (9.6)} \\ 
  equations & & &\\
 \hline
\end{tabular}
\caption{Accuracy of the different approximation schemes for the scan of Section \ref{sect:xSM}. All the numbers are percentages and are compared to the complete solution of the fluid equations including the Boltzmann equation of the top quark only. The numbers in parentheses were computed with a tanh {ansatz}.}
\label{table:accuracy}
\end{table}

\section{Summary and conclusions}
\label{sect:concl}

In this work we have rederived the fluid equations used to determine the wall velocity of a first-order phase transition bubble. Contrary to the standard methodology, we exactly solved the equations for the background temperature and fluid velocity, without linearizing in these quantities. This ensures the existence of nonsingular solutions regardless of the wall velocity, in particular for $v_w\gtrsim c_s$. We thereby overcome an important limitation of the standard formalism introduced in Ref.\ \cite{Moore:1995si}, which prevented the accurate description of supersonic walls. 

Furthermore, the Lorentz invariant fluid equations are derived from first principles;  EMT conservation is used to calculate the background temperature, fluid velocity and scalar field profiles, and a set of Boltzmann equations determines the deviation from thermal equilibrium of each species in the plasma. We proposed a spectral method to solve the Boltzmann equations, which has the attractive features of being simple, efficient, rapidly converging, and accurate. It allows one to 
approximately solve the Boltzmann equation directly with arbitrary accuracy, requiring just a single matrix inversion.

To explore the consequences of this new formalism on the dynamics of the bubble wall, we applied it to the singlet scalar extension of the SM. A key finding is that all bubble walls fall into one of two qualitatively distinct groups: either  deflagrations with $c_s\lesssim v_w\leq v_J$ (the Jouguet velocity), or ultrarelativistic detonations with $\gamma_w \gtrsim 10$. Hence, generic transitions will give bubble walls traveling either near the speed of sound, or that of light.  Although the result was derived in a specific model, we expect this classification to be a general feature of first-order phase transitions; it is a consequence of general equilibrium hydrodynamic effects. Since it mainly comes from equilibrium physics, departure from equilibrium has little impact on the terminal wall velocity. Rather than making a dominant contribution to the friction on the wall, as standard lore suggested, we find that it only gives a small correction, which changes the wall velocity by 2\% on average. However, in some borderline situations, it can generate a small additional pressure that transforms a would-be detonation into a deflagration; the out-of-equilibrium effects are significant in such cases.

Finally, we proposed several approximation schemes with increasing levels of complexity and accuracy. The simplest of these can, in most cases, correctly reproduce the qualitative behavior of the full treatment, while requiring almost no calculation. We also performed a numerical fit of the out-of-equilibrium pressure and pressure gradient from the present study. 
This allows future practitioners to quantitatively estimate the effects of friction, without having to solve the Boltzmann equations.  They can thereby improve the (already good) estimates from the local thermal equilibrium approximation,
with modest additional effort, to get 
accurate estimates of the wall speed.
It should be noted however that the methods discussed here do not address the question of how large $\gamma_w$ is in the case of highly relativistic walls.

A natural next step would be to extend the fluid equations derived in this work to $CP$-odd perturbations. This would be relevant to the study of electroweak baryogenesis, which is one of the 
primary applications of a first-order electroweak phase transition. Although generalizing the hydrodynamic and Boltzmann equations is straightforward,  the calculation of the collision integrals presents a challenge. In particular, some of the processes relevant for the $CP$-odd perturbations (such as those involving weak and strong sphalerons) should properly be calculated using lattice gauge theory, since they require
matrix elements that are not simply related to the total sphaleron rate. 
This is a problem we hope to investigate in future work.

\bigskip
{\bf Acknowledgements.}  We thank K.\ Kainulainen and T.\ Konstandin for stimulating discussions. This work was supported by the Natural Sciences and Engineering Research Council (NSERC) of Canada and the Fonds de recherche du Qu\'ebec Nature et technologies (FRQNT).
\begin{appendix}

\section{Collision integrals}
\label{app:coll}

We discuss here the calculation of the collision integrals by adapting the method used in Ref.\ \cite{Laurent:2020gpg} to be used with the spectral method discussed in Section \ref{sect:spectral}. The collision term for a given particle species is
\bea\label{eq:coll}
   \mathcal{C}[f] &=& \sum_i \frac{1}{2N_p} \int\frac{d^3k\, d^3p'\, d^3k'}{(2\pi)^5 2E_k 2E_{p'} 2E_{k'}}\,\vert\mathcal{M}_i\vert^2\nn\\
   &\times& \delta^4(p+k-p'-k')\,\mathcal{P}[f]\,;\\
  \mathcal{P}[f(p)] &=& f(p)f(k)\big(1\pm f(p')\big)\big(1\pm f(k')\big)\nn\\ &-&f(p')f(k')\big(1\pm f(p)\big)\big(1\pm f(k)\big)\,,
\eea
where the sum is over all the relevant processes. $p$ is the momentum of the incoming particle whose distribution is being computed, $N_p$ is its number of degrees of freedom,
$k$ is the momentum of the other incoming particle, and $p'$, $k'$ are the momenta of the outgoing particles. $|\mathcal{M}_i|^2$ is the squared scattering amplitude, summed over the helicities and colors of all the external particles. Finally, the $\pm$ appearing in population factor $\mathcal{P}$ is $+$ for bosons and $-$ for fermions.

To make the Boltzmann equation numerically tractable, $\mathcal{P}$ can be simplified by expanding it to linear order in the perturbations. Using the definition (\ref{eq:distrFunc}) of the distribution function and conservation of energy, one can show that $\mathcal{P}$ becomes
\bea
\label{eq:linearPop}
    \mathcal{P}[f] = f_p f_k f_{p'} f_{k'}&&\lp \frac{e^{\mathcal{E}_{\! p'}^\plasma/T}}{f_p^2}\delta\! f_p + \frac{e^{\mathcal{E}_{\! p}^\plasma/T}}{f_{p'}^2}\delta\! f_{p'}\right. \nn\\
    &&\ \ \left. -\frac{e^{\mathcal{E}_{\! k'}^\plasma/T}}{f_k^2}\delta\! f_k - \frac{e^{\mathcal{E}_{\! k}^\plasma/T}}{f_{k'}^2}\delta\! f_{k'} \rp,\qquad
\eea
where $f_p\equiv f_\eq(p)$.

Following the treatment of ref.\ \cite{Moore:1995si}, the calculation of the collision rates has been done to leading log accuracy, where it is justified to neglect the masses of all the
external particles, which implies $E_p=p$. One can also neglect $s$-channel contributions and the interference between diagrams because they are not logarithmic. To account for thermal effects, we use propagators of the form $1/(t-m^2)$ or $1/(u-m^2)$, where $m$ is the exchanged particle's thermal mass. It is given by $m_g^2=2g_s^2 T^2$ for gluons and  $m_q^2=g_s^2 T^2/6$ for quarks \cite{Weldon:1982bn}.

As previously discussed, we only consider the collision terms of the top quark. They are dominated by their strong interactions; we include only contributions to
$|{\cal M}|^2$ of order $g_s^4$. There are 3 relevant processes: top annihilation into gluons\footnote{As pointed out in Ref.\ \cite{Arnold:2000}, Ref.\ \cite{Moore:1995si} inadvertently omitted a $1/2$ symmetry factor in this amplitude.} ${\bar t}t\rightarrow gg$ and the two scattering $tg\rightarrow tg$ and $tq\rightarrow tq$. Their respective amplitudes are \cite{Moore:1995si}
\bea
\frac{1}{N_t}|\mathcal{M}_{tt\rightarrow gg}|^2 &=& -\frac{64}{9}g_s^4\frac{st}{(t-m_q^2)^2} \nn\\
\frac{1}{N_t}|\mathcal{M}_{tg\rightarrow tg}|^2 &=& -\frac{64}{9}g_s^4\frac{su}{(u-m_q^2)^2} + 16g_s^4\frac{s^2+u^2}{(t-m_g^2)^2} \nn\\
\frac{1}{N_t}|\mathcal{M}_{tq\rightarrow tq}|^2 &=& \frac{80}{3}g_s^4\frac{s^2+u^2}{(t-m_g^2)^2}.
\eea

To evaluate the integrals in (\ref{eq:coll}), one can first use the delta function and the symmetry of the integrand to  analytically
perform five of the nine integrals. The remaining four integrals can be evaluated numerically with a Monte Carlo algorithm like VEGAS.

\section{Boundary conditions of the hydrodynamic equations}
\label{app:BC}

For the convenience of the reader, we here reproduce the method of Ref.\ \cite{Espinosa:2010hh} to calculate the boundary conditions of the hydrodynamic equations (\ref{eq:EMT}). The quantities of interest are the temperatures $T_\pm\equiv T(\pm\infty)$ and the velocities of the plasma measured in the wall frame $v_\pm\equiv v(\pm\infty)$. 

By integrating the conservation equation for the energy-momentum tensor across the wall, one can show that the quantities $T_\pm$ and $v_\pm$ are related by the equations
\bea \label{eq:fluidEq}
v_+ v_- &=& \frac{1-(1-3\alpha_+)r}{3-3(1+\alpha_+)r}\,, \nn\\
\frac{v_+}{v_-} &=& \frac{3+(1-3\alpha_+)r}{1+3(1+\alpha_+)r}\,,
\eea
where $\alpha_+$ and $r$ are defined as
\bea \label{eq:EOS}
\alpha_+ &\equiv& \frac{\epsilon_+ - \epsilon_-}{a_+ T_+^4}\,, 
\quad
r \equiv \frac{a_+ T_+^4}{a_- T_-^4}\,, \nn\\
a_\pm &\equiv& -\frac{3}{4 T_\pm^3}\left.\frac{\partial V_\mathrm{eff}}{\partial T}\right\vert_\pm\,, \nn\\
\epsilon_\pm &\equiv& \left.\left(-\frac{T_\pm}{4}\frac{\partial V_\mathrm{eff}}{\partial T} + V_\mathrm{eff}\right)\right\vert_\pm\,.
\eea
These quantities are often approximated by the so-called bag equation of state, which is given in Ref.\ \cite{Espinosa:2010hh}. This approximation is expected to hold when the masses of the plasma's degrees of freedom are very different from $T$, which is not necessarily true in the broken phase. Therefore, we keep the full relations (\ref{eq:EOS}) in our calculations.

Subsonic walls always come with a shock wave preceding the phase transition front. Eqs.\ (\ref{eq:fluidEq}) can be used to relate $T_\pm$ and $v_\pm$ at the wall and the shock wave, but we need to understand how $T$ and $v$ evolve between these two regions. Assuming a spherical bubble and a thin wall, one can derive from the conservation of $T^{\mu\nu}$ the differential equations
\bea \label{eq:diffFluidEq}
2\frac{v}{\xi} &=& \gamma^2(1-v\xi)\left(\frac{\mu^2}{c_s^2}-1\right)\partial_\xi v\,,\nn \\
\partial_\xi T &=& T\gamma^2\mu\,\partial_\xi v\,,
\eea
where $v$ is the fluid velocity in the frame of the bubble's center and $\xi=r/t$ is the independent variable (not to be confused with $\xi$ in Eq.\ (\ref{eq:derivative})), with $r$ being the distance from the bubble center and $t$ the time since the bubble nucleation. In these coordinates, the wall is located at $\xi=v_w$. $\mu$ is the Lorentz-transformed fluid velocity
\be
\mu(\xi,v) = \frac{\xi-v}{1-\xi v}\,,
\ee
and $c_s$ is the speed of sound in the plasma
\be
c_s^2 = \frac{\partial V_\mathrm{eff}/\partial T}{T\,\partial^2 V_\mathrm{eff}/\partial T^2} \approx \frac{1}{3}\,.
\ee
The last approximation is valid for relativistic fluids, which is applicable in the unbroken phase. In the broken phase, some particles get a mass that can be of the same order as the temperature, somewhat reducing the speed of sound for those species.

One can find three different types of solutions for the fluid's velocity profile: deflagrations ($v_w<c_s^-$) have a shock wave propagating in front of the wall, detonations ($v_w>v_J$) have a rarefaction wave behind it, and hybrid transitions ($c_s^-<v_w<v_J$) have both shock and rarefaction waves. $v_J$ is the model-dependent Jouguet velocity, which is defined as the smallest velocity a detonation solution can have. Each type of wall has different boundary conditions that determine the characteristics of the solution. Detonation walls are supersonic solutions where the fluid in front of the wall is unperturbed. Therefore, it satisfies the boundary conditions $v_+=v_w$ and $T_+=T_n$. For that type of solution, Eqs.\ (\ref{eq:fluidEq}) can be solved directly for $v_-$ and $T_-$.

Subsonic walls always have a deflagration solution with a shock wave at position $\xi_{\rm sh}$ that satisfies $v_{\rm sh}^-\xi_{\rm sh} = (c_s^+)^2$, where $v_{\rm sh}^-$ is the fluid's velocity just behind the shock wave, measured in the shock wave's frame. It also satisfies the boundary conditions $v_-=v_w$ and $T_{\rm sh}^+=T_n$. Because these boundary conditions are given at two different points, the solution of this system can be somewhat more involved than for the detonation case. We use a shooting method, which consists of making a guess for $T_-$, solving Eqs.\ (\ref{eq:fluidEq}) for $T_+$ and $v_+$, integrating Eqs.\ (\ref{eq:diffFluidEq}) with the initial values $T(v_w)=T_+$ and $v(v_w)=\mu(v_w,v_+)$ for $\xi$ up to the point where  $\mu(\xi,v(\xi))\,\xi=(c_s^+)^2$ is satisfied. This procedure is iterated with different values of $T_-$ until Eqs.\ (\ref{eq:fluidEq}) are satisfied at the shock wave. Hybrid walls are characterized by $v_+ < c_s^- < v_w$,  with boundary conditions $v_- = c_s^-$ and $T_{\rm sh}^+ = T_n$, which make them similar to deflagrations.

\end{appendix}

\bibliography{ref}
\bibliographystyle{utphys}

\end{document}